\documentclass[traditabstract]{aa}
\usepackage{txfonts} 
\usepackage{graphicx}

\usepackage{txfonts}

\def\la{\raise.5ex\hbox{$<$}\kern-.8em\lower 1mm\hbox{$\sim$}}
\def\ma{\raise.5ex\hbox{$>$}\kern-.8em\lower 1mm\hbox{$\sim$}}

\def\kms{$\rm km\, s^{-1}$}
\def\cm3{$\rm cm^{-3}$}
\def\Ts{$\rm T_{*}$~}
\def\Vs{$V_{\rm s}$~}
\def\n0{$n_{\rm 0}$}
\def\B0{$B_{0}$}
\def\ne{$n_{\rm e}$~}

\def\Te{$\rm T_{e}$~}

\def\erg{$\rm erg\, cm^{-2}\, s^{-1}$}

\def\L12{L$_{12\mu m}$~}
\def\F12{F$_{12\mu m}$~}

\def\Hb{H$\beta$~}
\def\Ha{H$\alpha$~}
\def\Haa{H${\alpha}_{calc}$~}
\def\Ly{Ly$\alpha$~}

\begin{document}
  \title{Modelling  galaxy spectra at redshifts 0.2$\leq$z$\leq$ 2.3\\
by the  [OII]/\Hb and [OIII]/\Hb line ratios}

   \author{M. Contini \inst{1,2}
}

   \institute{Dipartimento di Fisica e Astronomia, University of Padova, 
              Vicolo dell'Osservatorio 2, I-35122 Padova, Italy 
                \and  
             School of Physics and Astronomy, Tel-Aviv University,
              Tel-Aviv 69978, Israel\\ 
             } 

   \date{Received }

\abstract
{We present the detailed modelling of line spectra emitted from galaxies at redshifts
 0.2$\leq$ z$\leq$ 2.3. The spectra account only for  a few oxygen  to \Hb line ratios.
The results show that   [OII]3727+3729/\Hb and [OIII]5007+4959/\Hb  
are not sufficient to  constrain the models. The data  at least of  an auroral  line,
e.g. [OIII]4363, should be known.
We have found  by modelling the  spectra  observed from ultrastrong emission line galaxy
and faint galaxy  samples, 
O/H relative abundances  ranging between 1.8$\times$10$^{-4}$ and 6.6$\times$10$^{-4}$.}


\titlerunning{Modelling galaxies at 0.2$<$z$<$2.3}
\authorrunning{M. Contini}

\maketitle
\section{Introduction}

Observations  of line spectra emitted from galaxies at relatively high redshifts are now available
(see Contini 2013b, hereafter Paper I, and references therein).
Metallicities, in terms of the relative abundance of the  heavy elements to H,
have been  calculated by a detailed modelling of the line ratios,
 providing  some information
  about galaxy evolution, star formation rates,  luminosities etc.

The spectra are relatively poor in number of lines at high z because only the strongest  ones are
observable. The modelling procedure, therefore, becomes problematic  regarding model degeneracy.

Line ratios corresponding to different elements in a spectrum depend  on  the physical parameters and on
the  relative  abundances of the elements.
The O/H ratio shows generally the highest relative abundance  compared with that of the other  heavy elements  whose
 lines are observed 
in the UV-optical -IR frequency range. 
Different solar O/H  are reported by Asplund et al (2009), Allen (1976) and Anders \& Grevesse (1989), namely, 
O/H=4.9 10$^{-4}$,  6.6 10$^{-4}$ and 8.5 10$^{-4}$, respectively.  
 Table 1 shows that Allen (1976)  relative abundance  values  range  between the two more 
recent results.
Therefore, we will refer to Allen (1976). Yet, the relative abundances are calculated consistently for each spectrum.
So the  values which appear in Table 1 are important only as references for  discussions.

\begin{table}
\caption{The solar element abundances}
\begin{tabular}{ccccccc} \hline  \hline
\ element & Allen & Anders \& Grevesse & Asplund et al.  \\ 
\         & (1976)       & (1989)      & (2009)      \\ \hline
\    H    &  12          &    12       & 12           \\
\  C      &  8.52        &  8.56       &  8.43       \\
\  N      & 7.96         &  8.05       &  7.83       \\
\  O      & 8.82         &  8.93       &  8.69  \\
\   Ne    & 8.            & 8.09       & 7.93 \\
\ Mg      & 7.4          &  7.58       &  7.6 \\
\ Si      & 7.52         &  7.55       & 7.51   \\
\ S       & 7.2          &  7.21       & 7.12   \\
\ Cl      &  5.6         &   5.5       & 5.5   \\
\ Ar      & 6.52         &   6.56      & 6.4   \\
\ Fe      & 7.5          &   7.67      & 7.5  \\ \hline\\
\end{tabular}
\end{table}
  
Line ratios  from the same element in different ionization stages e.g.
[OIII]5007+/[OII]3727+ (the + indicates that the $\lambda$$\lambda$5007,4959 and $\lambda$$\lambda$3727,3729 doublets are summed up), 
constrain the physical parameters such as the
photoionization flux reaching the gas, the temperature of the line emitting gas, etc .  
The oxygen line ratios to \Hb   constrain  the O/H  relative abundance. 
So the  "direct" or \Te ~ method (Seaton  1975, Pagel et al. 1992, etc) is  used to obtain  O/H from 
the observed oxygen  to \Hb line ratios.
By this method, the ranges of the  gas physical  conditions are   chosen among those   most  suitable to the
observed line ratios, e.g.
the  temperature   is calculated from  [OIII]5007+/[OIII] 4363,
 the  temperature and the density  ranges are  constrained by  [OII]3727+/\Hb  
considering, in particular,  that the critical density for collisional deexcitation 
of [OII] is $<$ 3000 \cm3.  The density is constrained also by
  [SII] $\lambda$6717/$\lambda$6731 line ratio, when observed.

In  Paper I we have modelled the spectra of galaxies   at redshifts between 0.001 and 3.4, 
on the basis of   at least hydrogen, oxygen and nitrogen lines. We have adopted a composed
model which accounts for the photoionization  and for shocks.

In the present  paper we  deal with spectra showing   a few oxygen  line ratios to \Hb, 
i.e. the [OIII] 5007+4959/\Hb, [OIII]4363/\Hb and [OII]3727+/\Hb observation data reported by Kakazu et al (2007)
for a sample of  {\it  ultrastrong emission line galaxies} at z$\sim$1   and the [OIII] 5007+4959/\Hb and 
[OII]3727+/\Hb observation data reported by Xia et al. (2012) for a sample of {\it faint galaxies} at 0.6$\leq$ z $\leq$ 2.4. 
Moreover, we have added a subsample of the stacking spectra of emission line galaxies
at 1.3$\leq$ z$\leq$2.3 reported by Henry et al (2013).
The stacking method is justified by the increasing number
 of  data observed in  the different galaxy samples.

Our aim is to discuss modelling degeneracy. We would like to find out the  smallest set of oxygen line ratios to \Hb  
suitable to constrain the O/H relative abundance.
Moreover, we will compare the O/H results  calculated by  detailed modelling  with those obtained by the 
 other methods, e.g. the direct method  and the metallicity calibrators (Perez-Montero \& Diaz 2005 and references therein).

We will model the spectra, even those showing a few lines, by the method adopted for  the spectra 
rich in number of lines (Paper I and references therein).
Namely, the gas is  ionized and heated by the black body radiation flux from the stars in the case of a starburst (SB) galaxy 
or by a power-law radiation flux from  the active nucleus in active galactic nuclei (AGN).

It  was found that galaxies at a relatively high z are the product of merging, 
therefore a shock dominated
regime is assumed leading to compression and heating of the emitting gas downstream of the shock front.
Collisional ionization and heating prevail on the radiation processes  at relatively high shock velocities.

When the observed line spectrum of a galaxy contains a few hydrogen lines, oxygen lines from no more than two
ionization stages, while  the lines from the other elements are missing,
 the results of the calculation process  can confront degeneracy. This refers at least  to the abundances 
of the elements that are relatively strong coolants, whose lines are not seen.  
Carbon lines are  not available in the optical range.
Neon is hardly included into dust grains and molecules due to its atomic structure, therefore 
neon could be useful to  investigate the evolution
of the heavy elements with z, but its lines are weak. Even if Ne/H $\geq$ N/H,  we will consider nitrogen
as the second  important heavy element because the N  lines (e.g. [NII] 6584, 6548) observed from luminous galaxies at 
redshifts as high as  z$\leq$3.5 (Paper I) are relatively strong.

We will check this issue  by the detailed modelling  of the  Kakazu et al (2007) sample  of ultrastrong line emission galaxies 
and the Xia et al. (2012) faint galaxy sample.

In this paper we will first model the spectra adopting a solar N/H  and discuss
degeneracy by reducing the N/H relative abundance.
In fact, the results obtained in Paper I
dealing  with spectra rich enough in number of lines,  show that N/H splits from $\geq$ 10$^{-4}$ to 
$<$ 10$^{-5}$  at  redshifts in the 0.2$\leq$ z $\leq$ 1 range.

The  observed spectra  are relatively  poor because  many significant lines
are missing (e.g. [NeIII], HeII, [OI], [NII] and [SII] etc) . Moreover,  the trend of  [OIII]5007+/\Hb versus  
[OII]3727+/\Hb  (Fig. 1) indicates that the distribution of the  
data follows the ionization parameter rather than the O/H  relative abundance, with some scattering
due to the different physical conditions in the different objects. 

To constrain the models by a first choice of the  physical  parameters, we have used
the grids (Contini \& Viegas 2001a,b) calculated previously by the  SUMA code for SBs and AGNs.
Then, we have refined the models in order to reproduce the observed line ratios.

In Sect. 2  the modelling method is presented. In Sect. 3 the Kakazu et al. (2007) sample is modelled. 
In Sect. 4 we  deal with the Xia et al. (2012)  spectra.
The results of Henry et al (2013) sample galaxies   appear in Sect. 5.
Discussion and concluding remarks follow in Sect. 6.

\begin{figure}
\includegraphics[width=9.0cm]{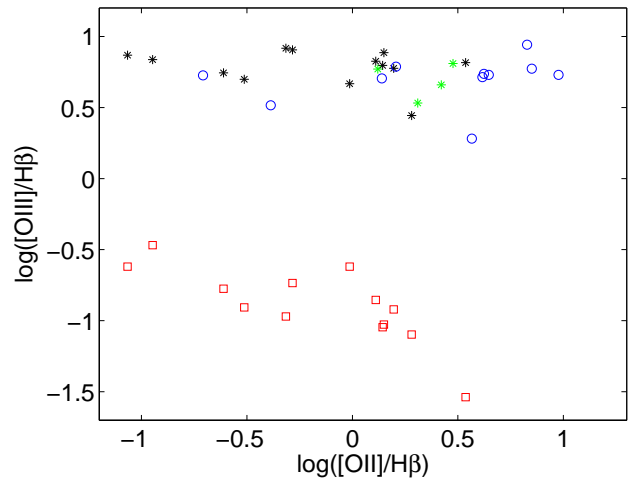}
\caption{The observed [OIII]5007+/\Hb versus [OII]3727+/\Hb 
by Kakazu et al : black asterisks; by Xia et al : blue circles: Henry et al. : green asterisks.
Red squares indicate the observed  [OIII]4363/\Hb versus [OII] 3727+/\Hb
line ratios by Kakazu et al.}
\end{figure}

\begin{figure}
\includegraphics[width=8.8cm]{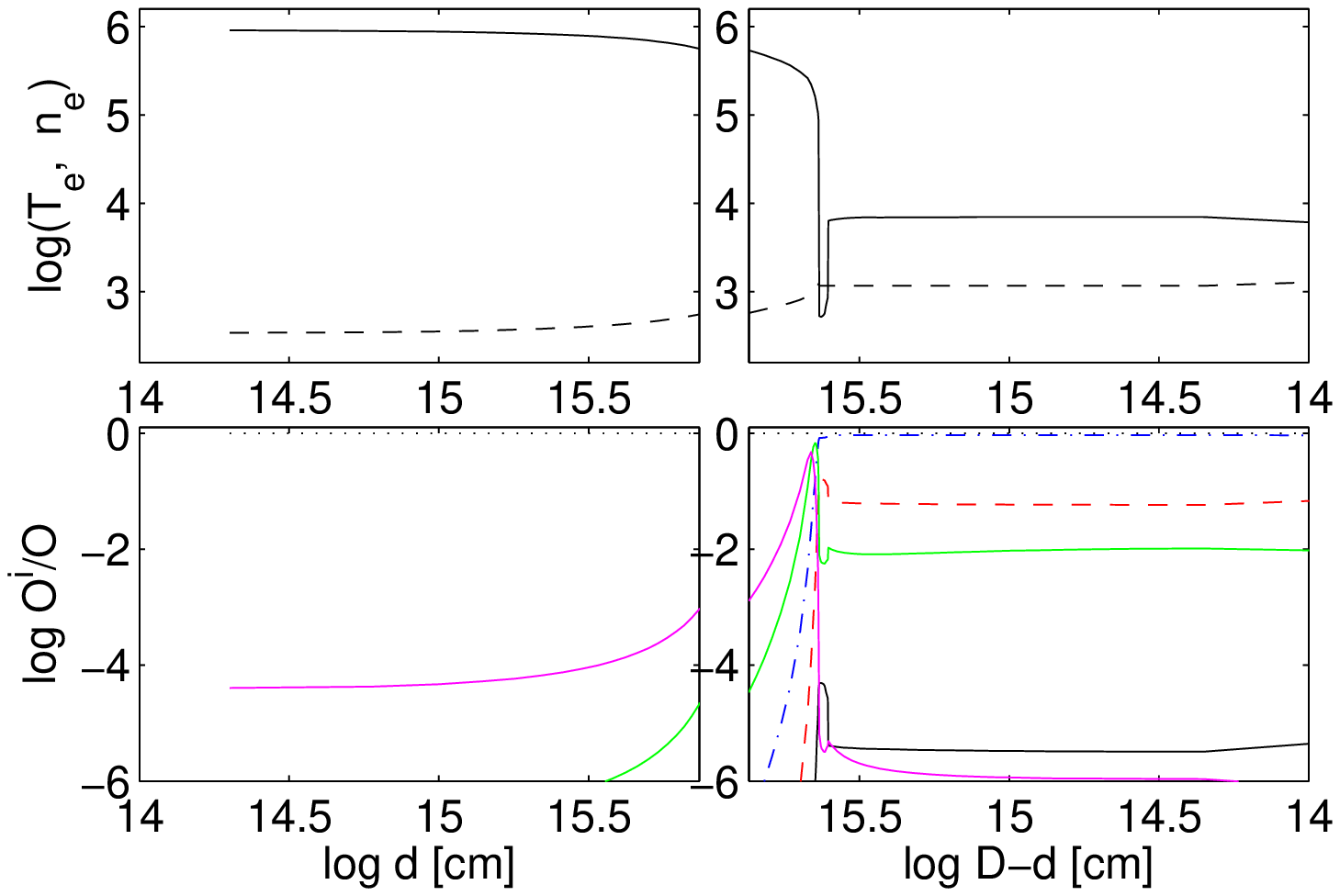}
\caption{Top panels : the profile of the electron temperature (solid line) and of
electron density (dashed line); bottom panels : the profile of  the  fractional oxygen ions :
O$^0$/O (black solid), O$^+$/O (red dashed), O$^{2+}$/O
(blue dot-dashed), O$^{3+}$/O (green solid), O$^{4+}$/O (magenta solid) and of H$^+$/H (red solid) throughout
a cloud corresponding to Kakazu et al ID 270 galaxy. The shock front is on the left of the left panel;
the photoionization flux reaches the right edge of the right panel.
The model corresponds to O/H=6.5 10$^{-4}$.
}
\includegraphics[width=8.8cm]{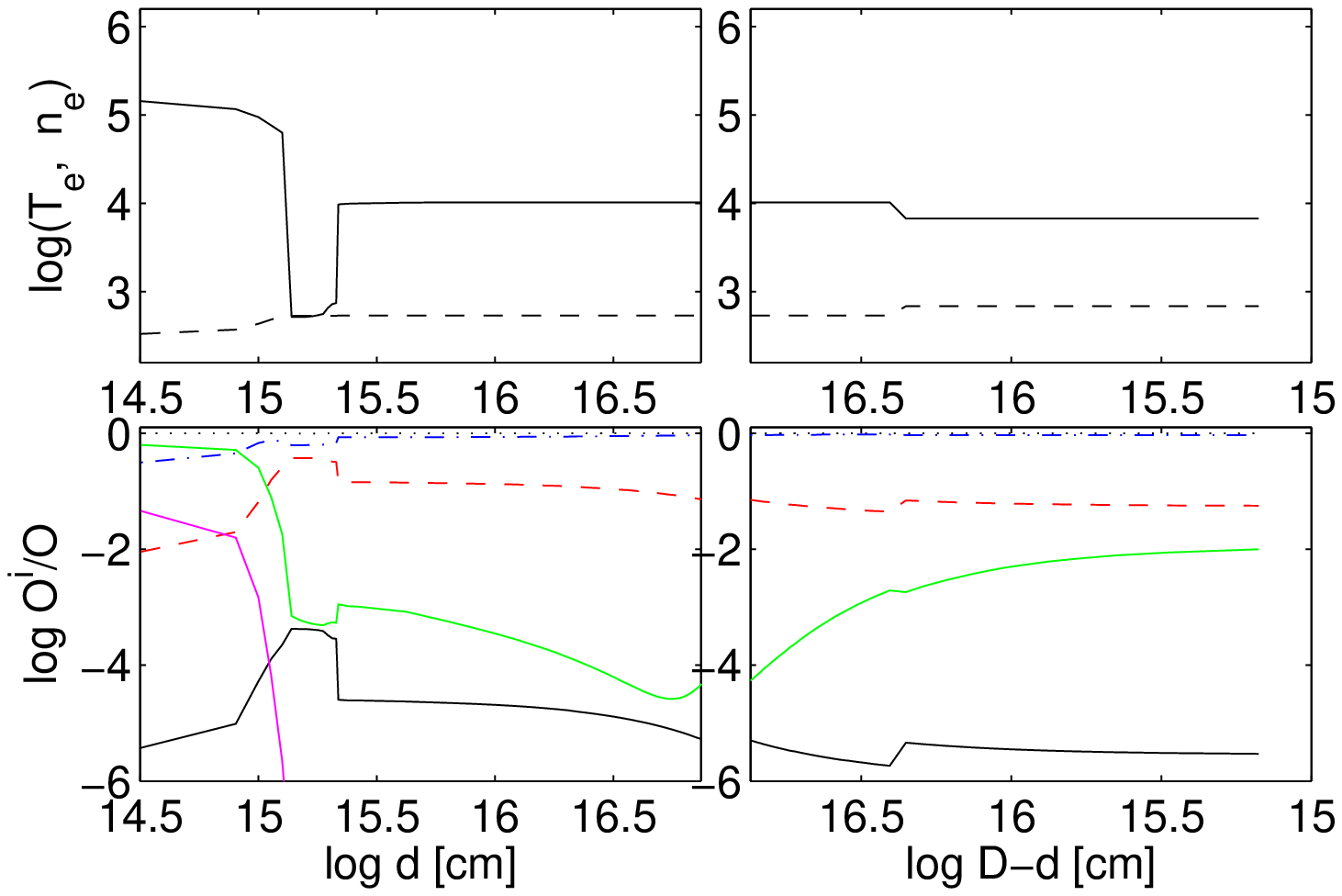}
\caption{The same as for Fig. 2 for models calculated by  O/H=1.8 10$^{-4}$.
}
\end{figure}

\section{Modelling method}

\subsection{Input parameters}
The  code
{\sc suma}\footnote{http://wise-obs.tau.ac.il/$\sim$marcel/suma/index.htm} is adopted for the calculation of the spectra,
because it   simulates the physical conditions in an emitting gaseous cloud under the coupled effect of
photoionization from an external radiation source and shocks. The line and continuum emissions
from the gas are calculated consistently with dust-reprocessed radiation in a plane-parallel geometry
(see Contini et al 2009 and references therein for a detailed description of the code).

 The input parameters are : the shock velocity \Vs, the preshock density \n0, the preshock 
magnetic filed \B0, {which depend on the shock.

The ionization parameter $U$ and the   effective star  temperature \Ts define the ionization flux
for starburst galaxies.
A  pure black body radiation referring to  \Ts  is a poor approximation  for a starburst, 
even adopting  a dominant spectral type (cf. Rigby \& Rieke 2004). However, the
line ratios  which are used to indicate \Ts also depend on
metallicity, electron temperature, density, ionization parameter, the
morphology of the ionized clouds  and, in particular, they depend on the hydrodynamical field.

The input parameter  that represents the radiation field  in AGNs is the power-law
flux  from the active  nucleus $F$  in number of photons cm$^{-2}$ s$^{-1}$ eV$^{-1}$ at the Lyman limit.
The spectral indices are $\alpha_{UV}$=-1.5
and $\alpha_X$=-0.7.

A magnetic field of 10$^{-4}$ gauss is adopted for all the models.
Moreover the relative abundances of the elements (He, C, N, O, Ne, Mg, Si, S, A, Cl, Fe) to H and the geometrical thickness 
of the emitting clouds $D$ are  important factors.

 The dust-to-gas ($d/g$) ratio is also an input parameter. It regards in particular the  infrared  frequency
range of 
the continuum spectral energy 
distribution (SED),  affecting the dust reprocessed radiation peak. The higher the intensity peak
relative to the gas bremsstrahlung, the higher the $d/g$ ratio is. Moreover, a high $d/g$ can reduce a non radiative
to a radiative shock (Contini 2004a) by  mutual heating and cooling of dust and gas. 
We have no data  for the continuum SED in the frequency range characteristic of the dust reradiation
peak for the galaxies in the present samples, therefore we cannot determine exactly the $d/g$ ratio. 
We adopted  an average $d/g$=0.003 which is  valid for starbursts.

     The ranges of the input parameters  are chosen with the following criteria.
 When the FWHM of the line profiles are not reported by the observations, we obtain a first hint 
about the shock velocity from the grids of models calculated for AGN and SB (Contini \& Viegas 2001a,b).
 \Vs affects in particular the [OII] 3727+3729/\Hb line ratio.
 The density of the emitting gas  could be deduced from the [SII]$\lambda$6717/$\lambda$6731 line ratios, 
when these lines are observed, because  the [OII]
doublet $\lambda$$\lambda$3727,3729 lines in galaxy spectra are generally blended.
The temperature of the starburst and the ionization parameter are determined directly from the fit
of the line ratios, in particular [OIII]5007+4959/\Hb and HeII/\Hb.

The geometrical thickness of the clouds ($D$) is a crucial parameter that can hardly be deduced from the observations.
Indeed, in the turbulent regime created by shocks, Rayleigh-Taylor and Kelvin-Helmholtz instabilities
cause fragmentation of matter that leads to clouds with very different geometrical  thickness coexisting in the same region.

\subsection{Modelling steps}

 The main differences between  the direct method and the  modelling by the code are 
described in the following.

The direct method  derives from the  oxygen lines (e.g. [OIII]5007 and [OII]3727)  
the physical conditions of the gas, in order to calculate  the element abundances. The
 temperature of the emitting gas is obtained by considering   the auroral line [OIII]4363.
In brief,  the direct method refers to the  temperatures and densities  most appropriated 
to  the observed line ratios.

By the code, the temperatures and the densities and the fractional abundances of the ions
that lead to specific line ratios, 
are all consistently calculated adopting a source of photoionization and heating of the gas  
(e.g. an SB or an AGN and/or shocks) throughout a cloud with certain characteristics.
Moreover,  the line intensities are all contemporarily  calculated {\it integrating} throughout the cloud, 
which is cut in a certain number of slabs  up to a maximum of 300.

The contribution of gas slabs in different conditions is the main cause of the different  
 relative abundances calculated
by the direct method and  by modelling.  In fact, the  line intensity increment
 corresponding to   a certain ion
calculated by the model, can be low in regions where the gas  conditions are less adapted.
The contribution of these regions  to the integration process leads to a weaker line.
Therefore, to   reproduce the observed line ratio to \Hb, a  relative  abundance
    higher  than that used by the direct method  is adopted.

Very schematically, by  modelling :

1)    we adopt an initial  input parameter set on the basis of the galaxy observations;

2)   calculate the density in the  slab of gas downstream from the compression equation;

3)   calculate the fractional abundances of the ions from each level for each element;

4)   calculate line emission, free-free  and free - bound emission;

5)   recalculate  the temperature of the gas in the slab by thermal balancing or the enthalpy equation;

6)   calculate the optical depth of the slab and the primary and secondary fluxes;

7)   adopt the parameters found in slab i as initial conditions for slab i+1.

8)  Integrating  on the contribution of the line intensities calculated in each slab, we obtain the
absolute fluxes of each of the lines, calculated at the nebula (the same for bremsstrahlung).

9) We then calculate the line ratios to a certain line (in the present case \Hb)

10)  and compare them with the observed line ratios.

The observed data have errors, both random and systematic. Models are generally allowed to reproduce the data
within a factor of 2. This leads to input parameter ranges of a few per cent. The uncertainty in the calculation
results (within 10 \%)  derives from the use of many atomic parameters, such as
recombination coefficients, collision strengths etc., which are continuously updated.
Moreover, the precision of the integrations depends on the computer efficiency.

Notice that the profiles of the ions follow the profiles of the physical parameters throughout a cloud
(as  can be seen in Figs. 2 and 3) ,
therefore each line ratio corresponds to a series of temperatures and densities of the emitting gas.

If the calculated spectrum does not fit the data, the calculations are restarted changing the input parameters.
We generally make a grid of models which is completed when the modelling results reproduce satisfactorily the data.
The role of the grid consists in showing which of the parameters are critical to the various line ratios.
By changing the input parameters (cross-checking all the line ratios for each model) the set which
reproduces the data can be selected.

The grid is calculated by the following steps :

 First we study in details all the characteristics of the galaxy in order to provide a first guess
of the input parameters.
 Then, we consider  the highest line ratio (generally, [OIII]5007/\Hb) and we test which of the input
parameters is the key to fit the [OIII]/\Hb line ratio.
 We try to  reproduce [OII]/\Hb and [NII/\Hb changing the physical parameters. By cross-checking the
[OIII]/\Hb ratio, all the process will be restarted many times until a fine tune of all the lines
is achieved. If some line ratios are not fitted, whichever the set of the  physical parameters, we
 change the relative abundances until all the data are 
 reproduced within about 20\% for the strong lines and  50\% for the weak lines.

\subsection{Relevant issues}

     From the modelling point of view, once the physical conditions of the emitting gas and the
photoionizing flux type and intensity are determined by the appropriated line ratios, the geometrical
thickness of the cloud is deduced from the ratio between relatively high and low ionization level lines.
In fact, larger clouds will contain a larger volume of gas at a relatively low temperature,
leading to stronger low ionization level lines. The choice of $D$, within the range of the observational
evidence, is then constrained by the best fit of  different line ratios.
The clouds are matter-bound or radiation-bound depending on the geometrical thickness as well as on 
$F$, $U$ and \Vs.

The coupled  effect of radiation and shock   determines the   electron temperature \Te and
  electron density \ne of the gas and  the  fractional abundances  $I_{i}/I$ of the ions  corresponding to
the  line  spectra. 
\Te, \ne and $I_{i}/I$ depend  on   the mutual heating (by radiation  and by collision)  and cooling (by recombination) 
throughout the gaseous clouds.
Therefore, the  emitting gas, even if it is  described by a single model referring to  one set of initial physical  parameters
and   element abundances, does not show  uniform  conditions.
For instance, the  temperatures  of the  emitting gas  range between 
T$_{max}(K)$ $\sim$ 1.45 10$^5$ (\Vs/ [100 \kms])$^2$ in the downstream region close to the shock front
and T$_{min}$ $<$ 10$^3$ K corresponding to the low ionization level and neutral gas. 
The gas  cools down and  recombines.
 The  cooling rate depends on free-free, free-bound radiation and line emission.
The line emission term  accounts for the gas composition, i.e. for the  abundances of all the
elements  composing the gas, even for those  elements whose lines  are not observed.
 The higher the  element abundances, the higher the  energy loss rates of the gas by
line emission. The  gradient of the temperature  drop downstream  close or far from  
the  shock front depends on \Vs, \n0 and  on the abundances of the elements.  
So a different O/H 
leads to  different  line ratios in general and  to different oxygen  to \Hb  line ratios, in particular.

When the gas clouds are ejected from the starburst, the shock front corresponds to the external edge
of the  cloud and the photoionizing (primary) flux from the stars reaches the opposite edge.
The line intensities are emitted from   the 
region downstream of the shock front, from the internal region of the cloud and from the region
facing the stars. These regions are bridged by the secondary diffuse radiation flux
which is calculated as well as the primary flux by radiation transfer throughout the nebula.
When the cloud is geometrical thin and the primary flux is strong, the cool region
inside the nebula is very reduced (see Figs. 2 and 3).

When both [OIII]/\Hb and [OII]/\Hb line ratios  overpredict the data, we tend to reduce  O/H.
However,  it is  evident that this will yield   different [OIII]/[OII] line ratios  which must be readjusted by
changing $F$ (or $U$ and \Ts in the SB case), \Vs, and/or $D$. The cross-checking process sometimes ends with unpredictable results
for O/H and the line ratios will be better reproduced by focusing on the physical parameters.

 The calculation of even a single line flux  needs  all  the physical and chemical parameters
 which appear for each model.

The absolute line fluxes referring to the ionization level i of element K are calculated by the
term n$_K$(i) which represents the density of the ion X(i). We consider that n$_K$(i) = X(i)[K/H]n$_H$, where X(i)
is the fractional abundance of the ion i calculated by the ionization equations, [K/H] is the relative abundance
of the element K to H and n$_H$ is the density of H (by number \cm3).
So the abundances of the elements are given relative to H as input parameters.
In models including the shock, compression (n$_H$/\n0) downstream  is calculated by the Rankine-Hugoniot
equations for the conservation of mass, momentum and energy throughout the shock front.

In this paper, we discuss the modelling of a galaxy on the basis of the  [OIII] 5007+/\Hb, [OII] 3727+/\Hb line ratios
 and eventually on  [OIII] 4363/\Hb.
Actually, the [OIII] 4363 line (generally weak and blended with H${\gamma}$) is not the  most suitable choice
 to constrain the models, but if we  aim to have some information for objects at high redshift,
 there is no alternative.

\section{The line ratios from the Kakazu et al. (2007) galaxy sample}

\begin{table*}
\centering
\caption{Modelling  the  line ratios (\Hb=1) observed by Kakazu et al (2007)}
\tiny{
\begin{tabular}{cccccccccccccccc} \hline  \hline
\  ID  &   z  & [OIII]5007+  &[OIII]4363  & [OII]3727+  & \Vs  & \n0   & \Ts       &  $U$  & O/H       & $D$   & \Hb abs  \\
\      &      &              &            &             & \kms & \cm3  & 10$^{4}$K &  -    & 10$^{-4}$ & cm    &   \erg \\ \hline
\ NB816&      &              &             &           &              &            &       &           &       &          \\
\ 40   &0.629 &  7.69$\pm$0.3&  0.094$\pm$0.034& 1.41$\pm$0.062&-  &  -    &  -        &  -    &  -        & -     &   -         \\
\ calc & -    & 7.3          & 0.1            & 1.3          & 200  &200    &5.         & 0.05   & 6.6      &5e15   & 0.005      \\
\ 76   &0.6319& 6.55$\pm$0.16&$<$0.029        & 3.44$\pm$0.08 &-   &  -    &     -      &  -    &  -        &  -    &   -        \\
\ calc &  -   &7.6           &0.022           & 3.5          & 200  & 70  & 2.2        & 7.6   & 6.6       &1.5e17 & 0.0056      \\
\ 118  &0.6439& 6.87$\pm$0.43&0.34$\pm$0.13   &0.113$\pm$0.026&-    & -   &     -      &  -    &   -       &  -    &    -        \\
\ calc &  -   & 7.           & 0.26           & 0.16         &250   & 70  & 4.4        & 1.    & 6.0       &1.5e16 & 2.9e-4       \\
\ 195  &0.628 & 4.65$\pm$0.03& 0.24$\pm$0.11 &0.97 $\pm$0.097&-    & -   &     -      &  -    &   -       &  -    &     -        \\
\ calc &   -  & 4.61         &0.15            & 0.7          & 230  &110  & 2.3        & 6.5   &3.         &4.e16  & 0.0056       \\
\ 252  &0.64  & 6.25$\pm$0.13&0.09$\pm$0.04   &1.39$\pm$0.037&-   &  -  &     -      &  -    &    -      &  -    &    -         \\
\ calc &  -   & 6.1          &0.094          & 1.1           & 200  & 200 & 5          & 0.05  & 6.5       & 5.e15 & 0.0064      \\

\ NB912&      &              &              &             &      &       &       &       &           &       &          \\
\ 3    &0.82  & 7.39$\pm$0.18& 0.24$\pm$0.08 &  0.086$\pm$0.025&-  & -    &  -   &    -      &   -   &   -       &  -    &   -      \\
\ calc &   -  & 7.5          &0.23           & 0.11        & 250  & 70   &   4.5     & 1.3   & 6.0       & 1.5e16& 2.8e-4    \\
\ 6    &0.83  & 8.04$\pm$0.49&0.184$\pm$0.11 &0.52$\pm$0.086&-   &  -   &   -       &  -    &   -       &   -   &   -       \\
\ calc &  -   & 9.5          &0.07           & 0.54         & 200  & 150  & 3.        & 1.7   & 4.2       & 6.e15 & 0.0013    \\
\ 9    &0.833 & 5.97$\pm$0.22& $<$0.12       &1.57$\pm$0.069& -    &  -   &   -       &  -    &   -       &   -   &    -      \\
\ calc &   -  & 6.           &0.04           & 1.6          & 150  & 80   &  2.5      & 2.     & 2.        & 1.e17 & 0.0029     \\
\ 10   &0.829 & 6.69$\pm$0.17&0.14$\pm$0.044 & 1.29$\pm$0.053&-   &   -  &   -       &  -    &   -       &   -   &   -       \\
\ calc &   -  & 7.0          &0.147          & 1.3          & 200  & 200  & 4.        & 0.12  & 6.6       & 4.e15 & 0.0034    \\
\ 20   &0.820 & 5.54$\pm$0.24&0.168$\pm$0.055&0.245$\pm$0.025&-    &  -   &  -        &  -    &   -       &   -   &   -       \\
\ calc &   -  & 5.4          &0.22           &0.2           &250  & 70   & 3.9       & 1.    & 6.0       & 1.5e16& 3.3e-4    \\
\ 239  &0.8273& 2.78$\pm$0.17&$<$ 0.08       &1.91$\pm$0.1  & -   &  -   &   -       &  -    &   -       &   -   &   -    \\
\ calc &  -   & 2.73         &0.072          &1.6           &200  & 200  & 3.8       & 0.05  & 6.5       & 5.e15 & 0.0067 \\
\ 270  &0.8176& 5. $\pm$0.23 &0.124$\pm$0.034&0.307$\pm$0.027&-  &  -   &  -        &  -    &   -       &  -    &   -    \\
\ calc &   -  & 5.           &0.11           &0.2           &250  & 80   & 3.8       & 0.8   & 6.5       & 1.5e16& 0.0011  \\
\ 60 $^1$ &0.393& 8.26$\pm$0.46&0.107$\pm$0.078&0.484$\pm$0.087&- &  -   &  -        &  -    &   -       &  -    &   -    \\
\ calc  &    -  & 8.29       &0.107          &0.68$\pm$     &200 & 200  & 5.        & 0.1   & 6.6       & 5.e15 & 0.0058  \\ \hline

\end{tabular}}

 $^1$ \Ha emitter

\end{table*}

\begin{figure}
\begin{center}
\includegraphics[width=9.6cm]{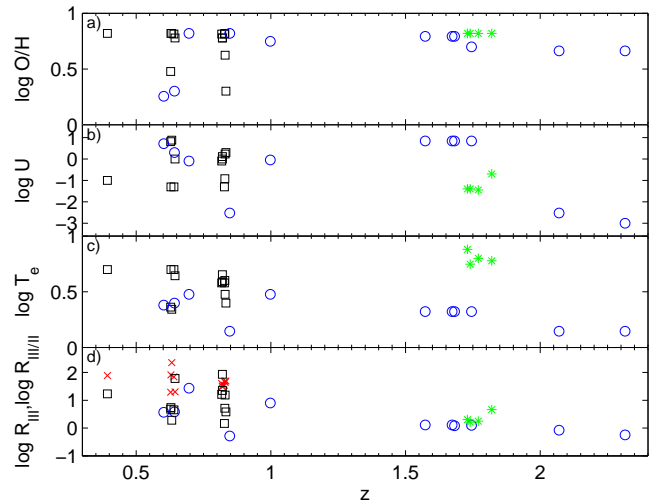}
\caption{Some  significant results calculated by modelling the Kakazu et al.  data (black squares),
 Xia et al. data (blue circles) and Henry et al data (green asterisks). 
In the bottom diagram R$_{III}$  indicates Kakazu et al [OIII]5007+/[OIII]4363 (red x);
R$_{III/II}$  indicates [OIII]5007+/[OII]3727+.}
\end{center}
\end{figure}

Narrowband observations by the 10m Keck II telescope 
of ultrastrong emission line galaxies were presented by Kakazu et al. 
The spectra refer to galaxies which show the [OIII] 5007+, [OIII] 4363, [OII] 3727+ and \Ha 
at redshifts z$<$ 1 and \Ly emitting galaxies at redshift z$>>$5. The final sample consists of 267 galaxies
 in the NB816 filter and 275 in the NB912 filter.
The ultrastrong emission line (USEL) galaxy data were obtained from intermediate
resolution (R$\sim$ 1000) spectroscopy for a subsample of 18 objects
that were specifically selected for a metallicity study by Kakazu et al.

We refer to the spectra from the objects selected by Kakazu et al and presented in their tables 4 and 5.
The observed  (reddening corrected) line ratios to \Hb=1 are  reported in Table 2
 in the rows showing the ID number of the galaxy. 
 In the next rows the calculated line ratios are given.
We have used the models referring to the starburst because 
 the  single galaxy luminosities are most probably  due to the starbursts and the
SFR is rather high  at  redshifts $\sim$ 0.6-0.8
corresponding to the observed spectra (Kakazu et al 2007).
In our sample we neglect the galaxies  where  [OII] 3727+ /\Hb   are missing or are
indicated  as upper limits.
In Table 2  the redshift is reported in column 2 and the line ratios appear in columns 3, 4 and 5. 

The errors in the observational data  include  uncertainties in the extinction and
the gaussian fit to the line profiles.
 The underlying stellar population  affects the continuum SED. Then, the relative intensity of the different lines
will be affected  when subtracting the underlying continuum  with different precision.
This  yields another  source of uncertainty in the results.

 The errors in the calculation results depend on the uncertainties of the various
coefficients and cross sections   and on the physical processes included in the models.

 In Table 2 columns  6-12 the input parameters  which yield the best fit to the data follow. In column 12 the 
absolute \Hb flux in  \erg calculated at the nebula is presented.
  A large gap (by a factor of $\sim$ 10$^{14}$)  
between the calculated and observed \Hb ($\sim$ 10$^{-17}$ \erg, Kakazu et al, fig. 7)
 can be noticed
because  \Hb is observed at Earth but calculated at the cloud.   The gap depends  on the square ratio of the
distance of the emitting cloud from the hot radiation source and the distance of the galaxy to Earth.

The results   presented in Table 2  show that
1) the ionization parameter is relatively high in the NB816
 galaxies with ID=76  and 195, indicating that
these objects are relatively compact, namely, the emitting clouds are close to the radiation source
and/or that the radiation
flux reaches the emitting  cloud undisturbed by  dusty and gaseous clumps in the interstellar medium.
For the NB912 galaxy sample at higher z (except ID 60 at z=0.393 in the bottom rows) the corresponding   
ionization parameters are lower. 

2) The temperature of the stars  in average is similar to  that calculated for SB galaxies in the same z range (Paper I).
The shock velocities are in the norm, and the preshock densities range between 70 and 200 \cm3.
  Relatively low \Ts, low \n0 and a high $U$ may suggest an  old age for these galaxies. 

3) The O/H relative abundances are solar in nearly all the objects with  minima of half and 0.3 
solar in NB912 ID= 195 and 9, respectively.
They result from  model calculations.

 The O/H calculated by Kakazu et al by the direct method are  lower than solar.  
 In fact, the lack of data
did not permit an accurate evaluation of  the  gas physical conditions.
Let us try to obtain  an acceptable fit to the observed line ratios adopting a model with a lower O/H.
We focus on the galaxy  ID 270 from the NB912 sample.
The observed line ratios are [OIII]5007+/\Hb=5.0, [OIII]4363/\Hb=0.124 and [OII] 3727+/\Hb=0.307 (Table 1, row 4 from bottom).
Adopting O/H =1.8 10$^{-4}$, \Vs=100 \kms , \n0=100 \cm3  $U$=0.4 and $D$=1.5 10$^{17}$ cm,
 we obtain [OIII[5007+/\Hb=5.1 and [OII]3727+/\Hb= 0.31, in  good agreement with the data. 
However, [OIII]4363/\Hb =0.03 is  low compared with the observed value (0.124).

Modelling the [OIII]4363/\Hb line ratios one should  take into consideration that the [OIII]4363 line is 
generally blended with
H${\gamma}$. The O$^{++}$/O fractional abundance   is at  maximum at  temperatures  
$\sim$ 10$^4$ - 10$^5$ K.
The  theoretical H${\gamma}$ /\Hb line ratios (Osterbrock 1989)  corresponding to these temperatures are  $\sim$ 0.45-0.46.
So  [OIII]4363/\Hb constrains the O/H  relative abundance (and the physical conditions)
in the galaxies, only if [OIII]4363 is uncontaminated from  blending and it is strong enough.
In this work we have constrained the models presented in Table 2  by the [OIII]4363/\Hb line ratios,
considering  that the data presented by Kakazu et al for [OIII] 4363 refer to this line only.

To better understand the line ratios emitted from the ID 270 galaxy we present in Figs. 2 and 3 the distribution 
of the electron temperature , the 
electron density and of the fractional abundance of the oxygen ions and of H$^+$/H throughout a cloud 
moving outwards from the SB.
The emitting cloud  is  divided into two halves represented by the left and right diagrams.
The left diagrams show
the region close to the shock front and the distance from the shock front on the X-axis scale is logarithmic. 
The right diagrams show the conditions
downstream far from the shock front, close to the edge reached by the photoionization flux which is opposite to the shock front. 
The distance from the
illuminated edge is given by a reverse logarithmic X-axis scale.
In  Fig. 2 the results refer to the model reported in Table 2, while in Fig. 3
the results refer to the model calculated by a relatively low O/H (1.8 10$^{-4}$).

Fig. 2 shows that the high temperature of the gas downstream ($>$ 10$^5$ K) depends on the shock velocity.
The temperature is $\sim$ 10$^4$ K  close to the cloud edge heated and ionized by the radiation flux from the star.
The  O$^{2+}$ and the H$^+$ ions  dominate a large region of the clouds. The [OIII]4363/[OIII]5007 line ratio
is higher in the model calculated by a higher \Vs because  the  temperature is high in a large zone of the 
cloud downstream.

\section{The  line ratios from  the Xia et al (2012) galaxy sample}

\begin{table}
\centering
\caption{The Extinction-corrected emission line fluxes of the PEARS/ERS Grism Galaxies}
\tiny{
\begin{tabular}{ccccccccc} \hline  \hline
\ ID&	z & 	[O II]3727$^1$  &          \Hb$^1$ &	[O III]5007$^1$ \\ \hline
\ 339&	0.602&	645.51$\pm$162.65&468.41$\pm$45.19 &2373.95$\pm$56.24\\ 
\ 364&	0.642&	80.90$\pm$15.93  &50.35$\pm$7.24  &308.61$\pm$9.59 \\
\ 246&	0.696&	4.50$\pm$4.50  &22.90$\pm$5.22 2  &21.91$\pm$6.92 \\	
\ 454&	0.847&	166.57$\pm$17.8&45.22$\pm$13.921&86.35$\pm$18.02 \\
\ 258&	0.998&	29.98$\pm$4.25&	73.63$\pm$35.74 &241.91$\pm$47.48 	\\
\ 432&	1.573&	101.97$\pm$23.&	24.21$\pm$11.76 &132.11$\pm$15.56 \\
\ 563&	1.673&	93.91$\pm$17.3&	13.95$\pm$9.06  &122.04$\pm$11.86\\
\ 103&	1.682&	43.55$\pm$10.2&	9.84$\pm$7.81   &2.83$\pm$10.33 \\
\ 195&	1.745&	87.84$\pm$13.87&21.25$\pm$8.28  &109.87$\pm$10.91\\	
\ 242&	2.070&	94.79$\pm$29.03&13.39$\pm$8.57  &79.46$\pm$11.19 \\
\ 578&	2.315&	116.58$\pm$21.06&12.29$\pm$10.42 &65.98$\pm$13.53 \\ \hline
\end{tabular}}

$^1$ in 10$^{-18}$ \erg

\end{table}

\begin{table*}
\centering
\caption{Modelling  the  line ratios (\Hb=1) observed by  Xia et al (2012)}
\tiny{
\begin{tabular}{cccccccccccccccc} \hline  \hline
\ ID   &   z  & [OIII]5007+& [OII]3727+    & \Vs & \n0   & \Ts       &  $U$  & O/H       & $D$   & \Hb abs & \Hb obs      \\
\      &      &             &           & \kms & \cm3 & 10$^{4}$K &       & 10$^{-4}$ & cm    & \erg & 10$^{-18}$ \erg \\ \hline
\ 339  & 0.602& 5.07       & 1.38      &   -  &   -  &   -       &  -    &           &  -    &    & 468.41 $\pm$ 45.19 \\
\ calc & -    & 5.4        &  1.3      & 200  & 80   & 2.4       & 5.2   & 1.8       & 1.e17 & 5.1e-3  & -  \\
\ 364  & 0.642& 6.13       & 1.61      &  -   & -    & -         & -     & -         & -     & -     &50.35 $\pm$7.24 \\
\ calc &  -   & 6.0        & 1.61      & 150  & 80   & 2.5       & 2.    & 2.        & 1.e17 & 2.9e-3&-    \\
\ 246  & 0.696& 5.32       & 0.196     & -    &-     &-          &-      & -         &   -   &   -  & 22.90$\pm$5.22     \\
\ calc &   -  & 5.0        & 0.2       & 250  & 80   & 3.        & 0.8   & 6.6       & 1.5e16& 1.13e-3&-   \\
\ 454  & 0.847& 1.91       &3.68      &   -  &  -   &  -        &  -    &  -        &  -    &   -   & 45.22$\pm$13.92   \\
\ calc &    - & 2.1        & 3.2      & 250  & 180  & 1.4       & 0.003 & 6.6       & 5.e15 & 2.6e-3 &-  \\
\ 258  & 0.998& 3.28       &0.41      & -    &  -   &  -        &  -    &  -        &  -    &  - &73.63$\pm$35.74$^*$   \\
\ calc &  -   & 3.66       & 0.44     & 280  & 100  & 3.        & 0.9   & 5.6       & 1.2e16& 1.5e-3 &-  \\
\ 432  & 1.573& 5.46       & 4.2      & -    &  -   &  -        &  -    &  -        &  -    &   -  &24.21$\pm$11.76$^*$ \\
\ calc &   -  & 5.5        & 4.4      & 200  & 70   & 2.1       & 7.    & 6.2       & 1.4e17& 5.3e-3&-   \\
\ 563  & 1.673& 8.75       & 6.73     & -    &  -   &  -        &  -    &  -        &  -    &  - &13.95$\pm$9.06$^*$     \\
\ calc &  -   & 8.63       & 6.1      & 200  & 70   &  2.1      & 7.    & 6.2       &1.3e17 & 3.8e-3&-  \\
\ 103  & 1.682& 5.37       &4.43     & -    &  -   &  -        &  -    &  -        &  -    &  - &9.84$\pm$7.81$^*$     \\
\ calc &   -  & 5.5        & 4.4      & 200  & 70   & 2.1       & 7.    & 6.2       & 1.4e17& 5.3e-3 &- \\
\ 195  & 1.745& 5.17       & 4.13     & -    &  -   &  -        &  -    &  -        &  -    &  - &21.25$\pm$8.28$^*$     \\
\ calc &   -  & 5.23       & 4.2      & 200  & 70   & 2.1       & 7.    & 5.        & 1.44e17&5.23e-3 &-\\
\ 242  & 2.070& 5.93       & 7.08    &  -   &  -   &  -        &  -    &  -        &  -    &   - & 13.39$\pm$8.57$^*$  \\
\ calc &   -  & 5.22       & 7.6     & 250  & 165  & 1.4       & 0.003 & 4.6       & 5.e15 & 7.6e-4 &- \\
\ 578  & 2.315& 5.37       & 9.48    &  -   &  -   &  -        &  -    &  -        &  -    &  - &12.29$\pm$10.42$^*$   \\
\ calc &   -  & 5.9        & 9.2     & 250  & 165  & 1.4       & 0.001 & 4.6       & 5.e15 & 6.7e-4 &- \\ \hline

\end{tabular}}

$^*$ The 3 $\sigma$ upper limit of the \Hb line is used for galaxies with S/N $<$ 3.

\end{table*}

\begin{table*}
\caption{Comparison of calculated line ratios to \Hb=1 with Henry et al data}
\tiny{
\begin{tabular}{lccccccccccccccc} \hline  \hline
ID     &   z  & [OII]3727+  &  [OIII]5007+  & \Ha & \Hb (abs) & \Vs  & \n0    & $F$      & \Ts    &$U$ & $D$          & O/H   \\
\       &   - &     -       &      -        &  -      & \erg      & \kms   & \cm3   & units$^1$&10$^4$ K& -  & 10$^{18}$ cm & 10$^{-4}
$ \\ \hline
\   1   & 1.82 & 1.32      & 5.89$\pm$1.23 &3.6      &  -        & -       &  -     &  -      &-  &-    & -        & -       \\
\ m1$_{SB}$ &-  & 1.3      & 6.            &2.94     & 0.093     & 180     & 200    & -       & 6 &0.2  & 5        & 6.6      \\
\ m1$_{AGN}$ & - & 1.58    & 5.5           &3.3      & 0.118     & 130     & 270    & 5       &-  &-    & 7        &  4.       \\
\   2   & 1.73 & 3.        & 6.45$\pm$1.2  &3.6      & -         &  -      &  -     &  -      &-  &-    & -        &   -     \\
\ m2$_{SB}$&-&  3.1        & 6.3           &2.93     & 0.045     & 240     & 200    & -       &7.6&0.04 & 9         & 6.6   \\
\ m2$_{AGN}$& -   & 2.9    & 6.5           &3.3      & 0.065     & 130     & 220    & 2.3     &-  &-    & 7        & 6.     \\
\ 3     & 1.77 & 2.64      & 4.57$\pm$1.15 &3.6      &  -        &  -      &  -     &   -     &-  &-    & -        &  -      \\
\ m3$_{SB}$& -  & 2.6      & 4.58          &2.94     & 0.043     & 180     & 220    & -       &6.3&0.036& 7         & 6.6   \\
\ m3$_{AGN}$&-   & 2.5     & 4.85          &3.44     & 0.079     &  160    & 260    & 2.1     &-  &-     & 7        &  6.6    \\
\ 4     & 1.74 &  2.04     & 3.4$\pm$1.17  &3.6      & -         &   -     &  -     &   -     &-  & -    & -        &  -     \\
\ m4$_{SB}$ & -&  2.1      & 3.4           &2.94     & 0.051     & 180     & 240    & -       &5.6& 0.04 & 5        & 6.6    \\
\ m4$_{AGN}$ & - & 2.04    & 3.37          &3.5      & 0.106     & 130     & 260    & 2.      &-  & -    & 9        & 6.0    \\ \hline
\end{tabular}}

$^1$  10$^{10}$ photons cm$^{-2}$ s$^{-1}$ eV$^{-1}$ at the Lyman limit

\end{table*}

Xia et al (2012) presented the spectra of faint galaxies at 0.6$<$z$<$2.4 observed by 
Advanced Camera for Surveys (ACS) on the {\it Hubble Space Telescope (HST)} 
and in the near-infrared using Wide-Field Camera 3. 
Xia et al data come from low resolution (R$\sim$ 100) grism spectroscopy in
which even the \Hb-[OIII] lines are heavily blended.

 The line flux  uncertainties are shown in Table 3.
The modelling results of the [OII] 3727+/\Hb and [OIII] 5007+/\Hb line ratios are  shown in Table 4.
Table 4 array is similar to that of Table 2.  
 In the last column of Table 4 the observed \Hb fluxes are given. 

The spectra presented by Xia et al show the minimum number of lines which can yield  reliable results.
In fact, we have run   a grid of many models ($\sim$ 30) for each spectrum before selecting the line ratios
best fitting the data.

The results for the whole sample show rather low \Ts ($<$ 3 10$^4$ K) and  $U$ ranging throughout four orders. 
The shock velocities and the preshock densities are in the ranges calculated for the Kakazu et al sample (Fig. 4),
lower than those for SB and AGN which are  shown in Fig. 5.
Actually in Fig. 5 we report the results presented in Paper I.

The O/H relative abundances are  about solar
at  0.696 $\leq$ z $\leq$ 1.745, 0.3 solar for galaxies at 0.602 and 0.642 and increase to $\sim$ 0.7
solar at z$>$2.

We have chosen the spectrum observed from the galaxy ID 195 (Table 4, row 6 from bottom) to investigate degeneracy.
The detailed modelling of the [OII]3727+/\Hb and [OIII]5007+/\Hb line ratios leads to O/H=5. 10$^{-4}$.
This corresponds to 12+log(O/H)=8.7, while  Xia et al. obtained  8.38 (-0.27)  adopting
metallicity calibrators.
We have tried to fit the oxygen line ratios by a lower O/H. We have found
 [OII]/\Hb=4.54 and [OIII]/\Hb=5.43 in very good agreement with the data (Table 4)
adopting O/H = 1. 10$^{-4}$ (12+log(O/H)=8.), \Vs=200 \kms, \n0=67 \cm3, \Ts=1.9 10$^4$, $U$=6.8, $D$=9.10$^{16}$ cm.
Indeed the  modelling of the Xia et al spectra shows degeneracy.
However,  the
 model presented for ID 195 in Table 4  yields [OIII]4363/\Hb = 0.017, while the
model calculated with a low O/H shows [OIII]4363/\Hb = 0.144. [OIII]4363/\Hb ratios  between 0.144
and  0.04 are reported  by  Kakazu et al.  (Table 2). So they are both reasonable.
In their paper, Xia et al. (2012)  mention the weakness of the [OIII] 4363 line in the observed spectra. 
This   information can be used to constrain the results,  and  it  may indicate
that the model  calculated with a higher O/H (as that presented in Table 4) is more reliable.

\section{Modelling the sample by Henry et al (2013)}

Henry et al (2013)  report mass-metallicity relation for log (M/M$_{\odot}$) between 8 and 10 calculated by
stacking spectra of 83 emission-line galaxies with redshifts 1.3 $\leq$ z $\leq$ 2.3.
In their table 1 they present line ratio observations for four stacked galaxies at  
1.74$\leq$z$\leq$ 1.82,
 covering
the ([OII]3727+3729 + [OIII]5007+4959)/\Hb line ratios which  are adapted to the metallicity diagnostic 
(Pagel et al.1979).
 In fact,  Henry et al used  the metallicity calibrator method.
 The sample is taken from Hubble Space Telescope Wide Field Camera 3 grism observations.

We  expand our investigation on the O/H relative abundances calculated for galaxies on
the basis of the [OII]/\Hb and [OIII]/\Hb lines  including the Henry et al.  (2013, table 1) observations.
The spectra are not specific to single galaxies and must be considered as averages within small ranges.
We refer to the  reddening corrected data.
Anyhow, the differences between the corrected and not corrected line ratios
are within the observed errors.

In Table 5 we  report the  observed line ratios and compare them with model calculations results.
The [OII] 3727+/\Hb ratios  were calculated from the Henry et al O32 ([OIII]5007+/[OII] 3727+)
reported in their table 1. However, the
 errors of the [OII] 3727+/\Hb line ratios  cannot be calculated because the observed \Hb fluxes are not 
given by Henry et al.  

The data for each group  of galaxies  are followed by the calculation  results  in the next
 two rows, one referring to models calculated for the SB and the next for the AGN. 
In fact, Henry et al mention an
eventual contribution to the SB  of  AGN spectra. 
 Notice that the data refer to averaged spectra for a group of heterogeneous galaxies therefore the results
of modelling should be considered only  as approximations.
The input parameters
adopted by the models  are presented in the last 7 columns of Table 5.

The results show that the O/H  referring to the ID= 1  spectrum calculated by the AGN  is 4.10$^{-4}$.
All the other  galaxies show solar O/H = 6.0 -  6.6 10$^{-4}$ (Allen et al 1976). 
Moreover, regarding the starburst,
the star temperatures are $\geq$ 5.6 10$^{4}$ K.  The radiation flux from the AGN components are
similar to  the lower limit  of $F$ for AGN  but higher than  the low luminosity AGNs  fluxes (e.g. Contini 2004b).

\section{Discussion and concluding remarks}

The spectra observed from the  galaxies presented    by Kakazu et al. and by Xia et al.  surveys are  employed
 to calculate the physical conditions of the emitting gas by a detailed modelling  of ultrastrong emitting line galaxies 
and faint galaxies, respectively, at intermediate redshifts.
Moreover, we have modelled the spectra presented by Henry et al which were obtained from the stacking
of 85 luminous galaxies.

The results of the most significant parameters are shown in Fig. 4. The O/H ratios are shown in units of 10$^{-4}$
and \Te in units of 10$^4$ K. All the results presented in Tables 2,  4 and 5  are compared with those calculated for
a much larger sample of galaxies in Fig. 5.

\begin{figure*}
\begin{center}
\includegraphics[width=7.8cm]{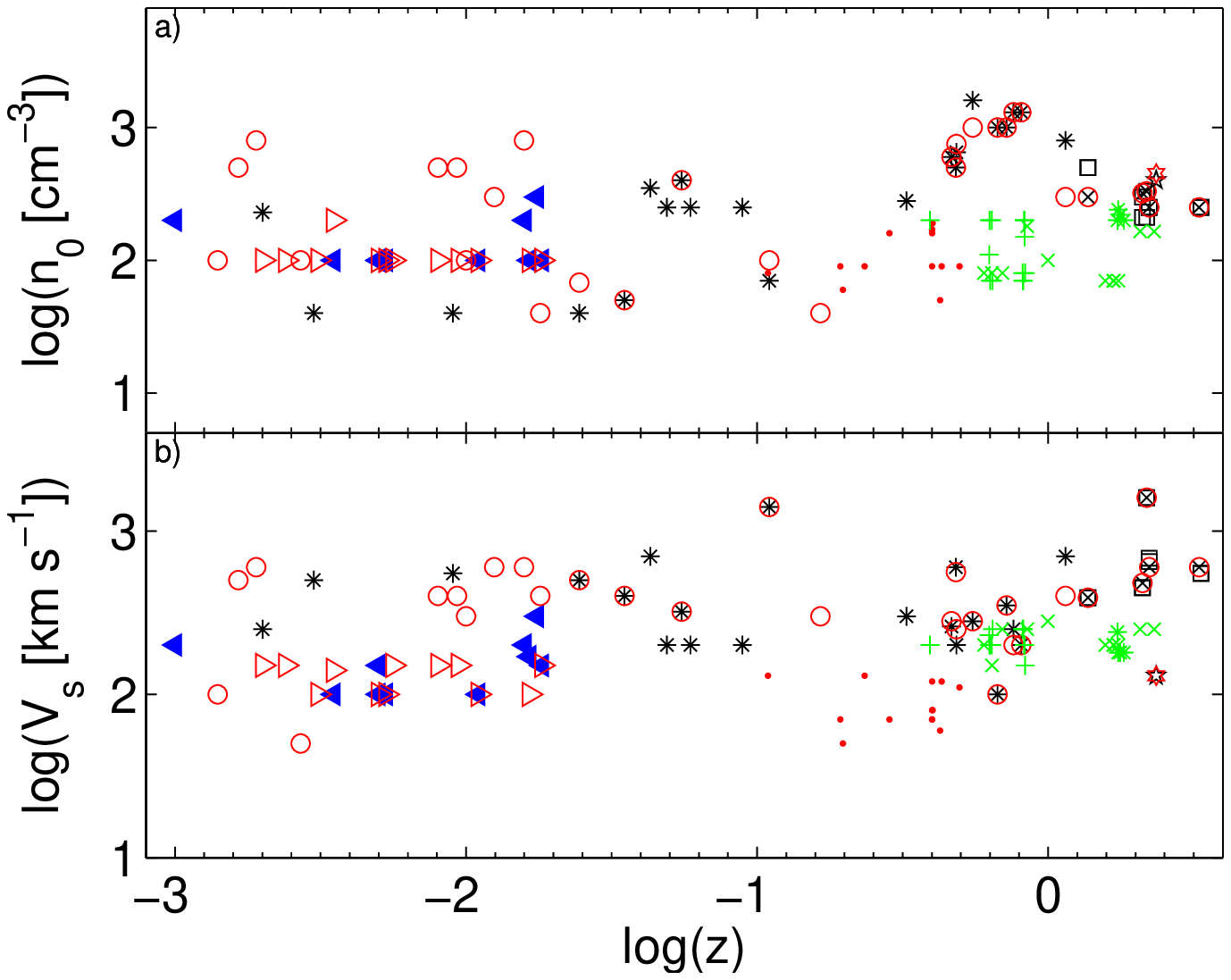}
\includegraphics[width=7.8cm]{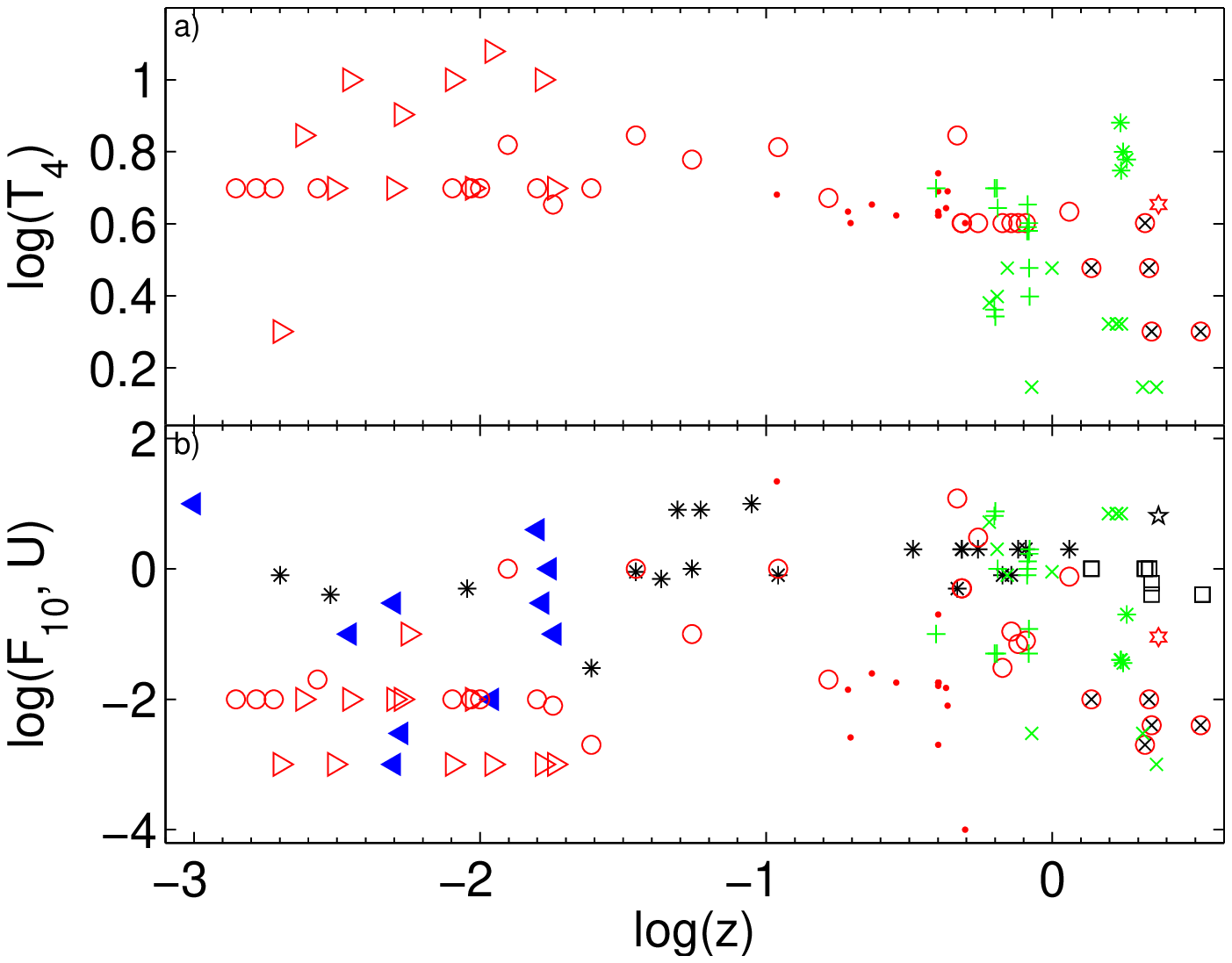}
\includegraphics[width=7.8cm]{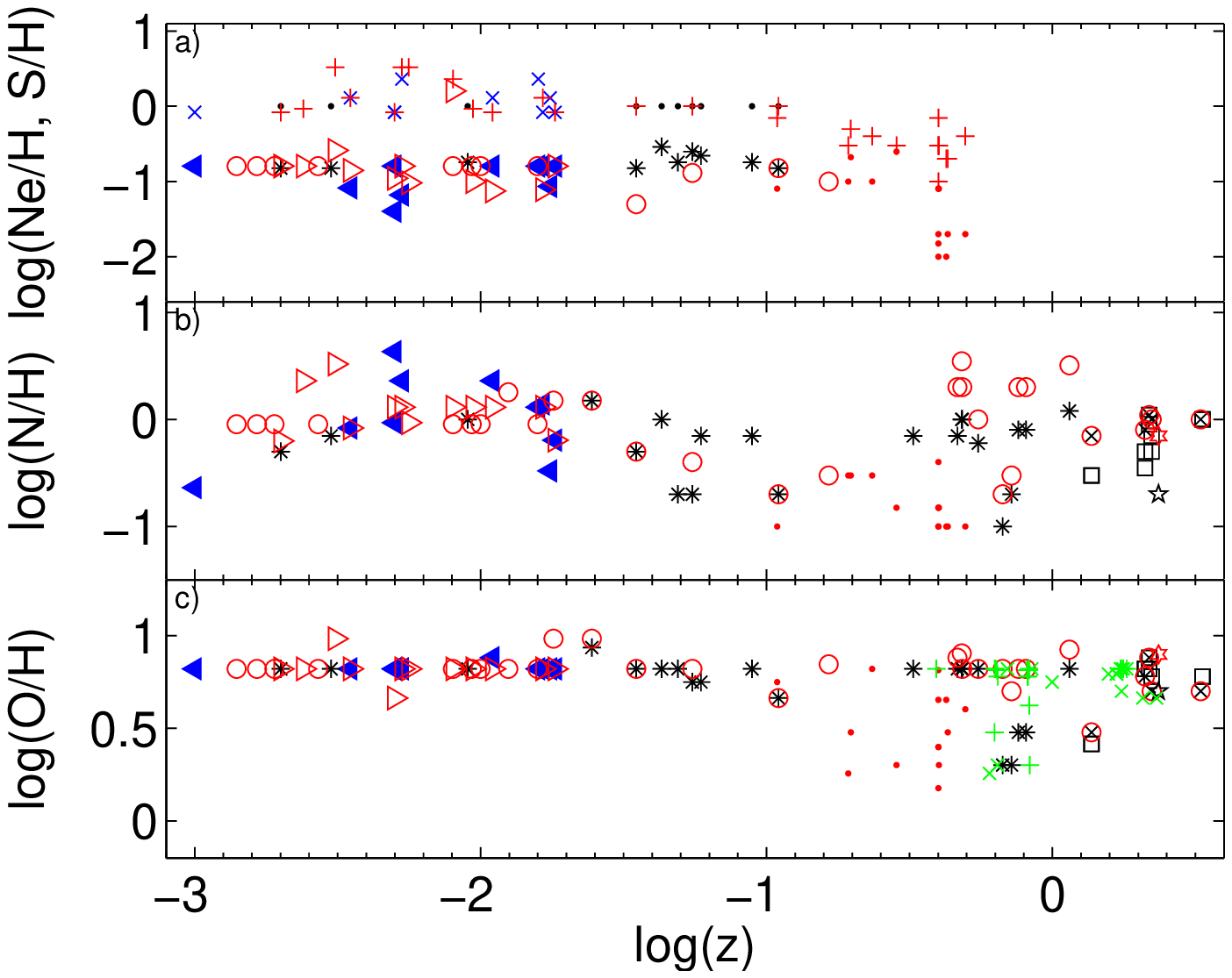}
\includegraphics[width=7.8cm]{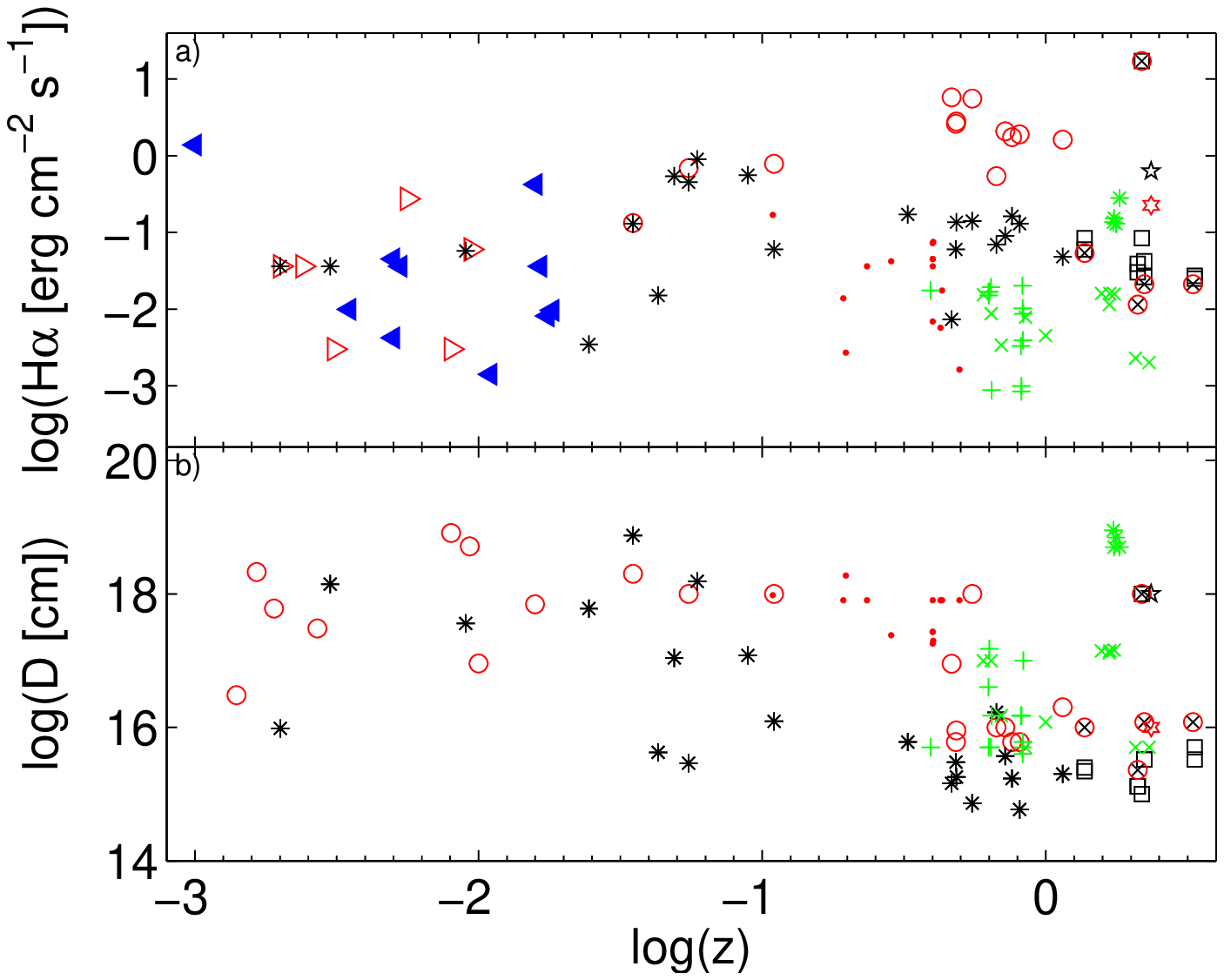}
\includegraphics[width=7.8cm]{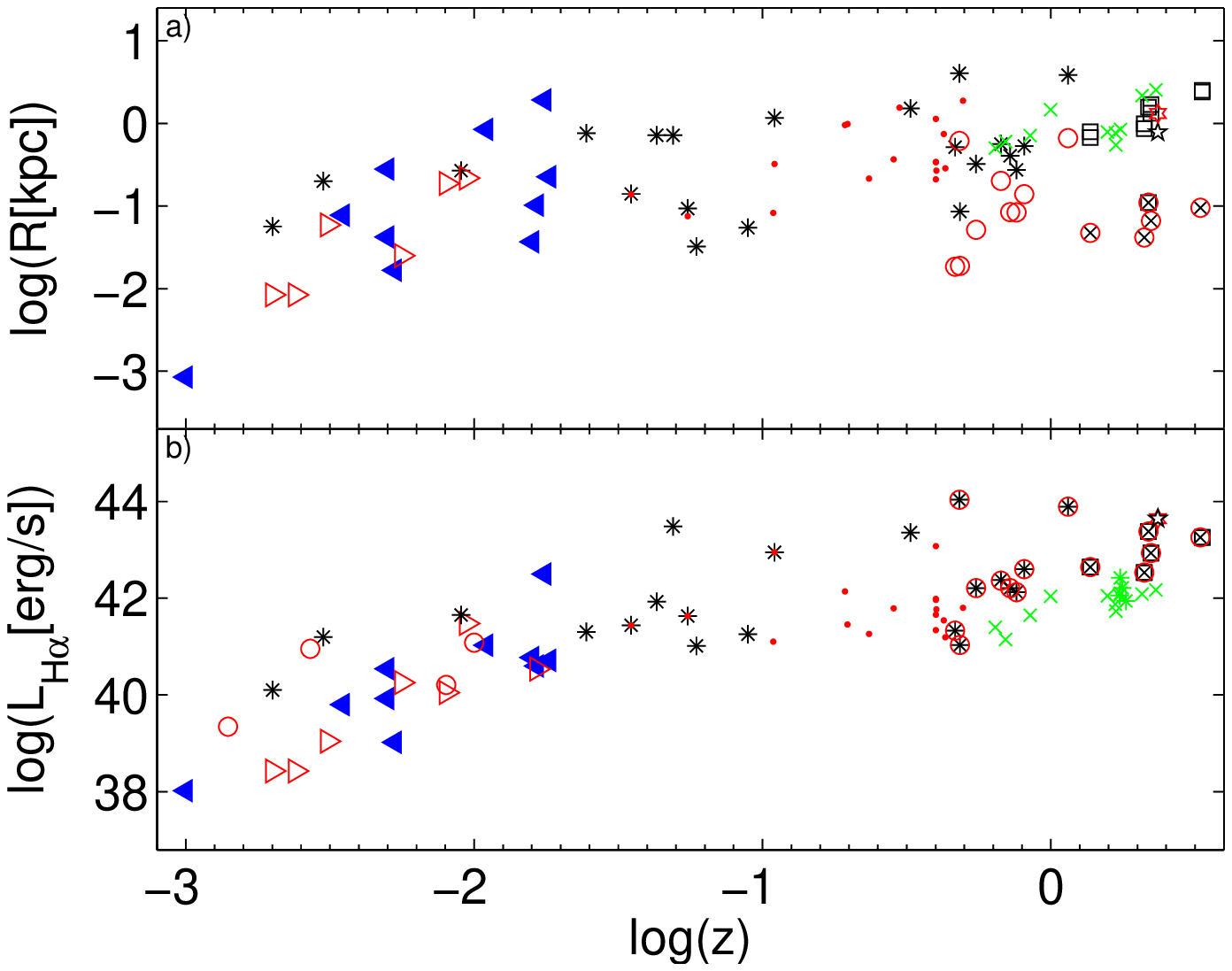}
\includegraphics[width=7.8cm]{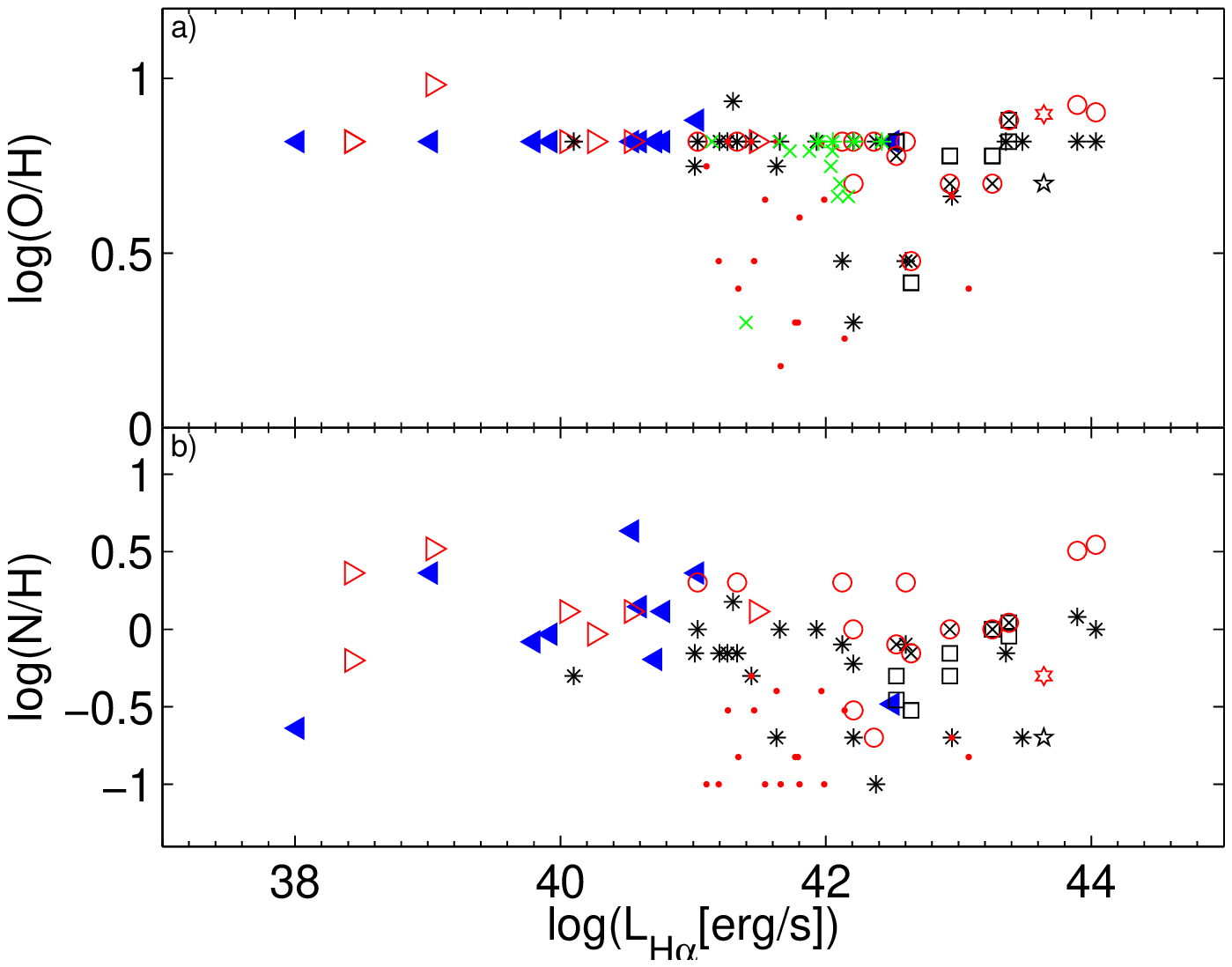}
\caption{The evolution of different parameters with z (adapted from Paper I, Fig. 3).
Top left : \n0 and \Vs; top right : the radiation parameters $F$ (in units of 10$^{10}$ 
photons cm$^{-2}$ s$^{-1}$ eV$^{-1}$ at
the Lyman limit), the ionization parameter $U$ and the temperature of the stars (in units of 10$^4$ K);
middle left : the relative abundances in units of 10$^{-4}$; middle right : \Haa and $D$; bottom  left:
the radius R (in kpc) of the NLR  in AGN  and of the emission regions in SBs
and the \Ha luminosity (L$_{H\alpha}$): bottom right : the metallicity versus L$_{H\alpha}$.
Symbols : green + , green x refer and green asterisks refer to the Kakazu et al,  Xia et al and
Henry et al samples, respectively.
Red circles represent SB galaxies
(Viegas et al. 1999, Contini 2013a,  Ramos Almeida et al 2013, Capetti et al 2013,
Winter et al. 2010);
 red circles encircling a x refer to the SBs in the optically faint ULIRGs (Brand et al. 2007);
red  triangles : SB  galaxies in the LINER sample
(Contini 1997);
red dots :  HII regions in star forming galaxies galaxies
(Kobulnicky \& Zaritsky 1999);
 red hexagram : ULIRG in QSO 2222-0964 (Krogager et al. 2013);
 black asterisks  : the AGNs
(Ramos Almeida et al. 2013, Contini 2013a, Schirmer et al. 2013, Winter et al. 2010);
 black squares refer to the optically faint ULIRGs (Brand et al. 2007);
 blue filled triangles : the AGNs belonging to the LINER sample (Contini 1997);
black pentagram  : ULIRGs (Krogager et al 2013).
 For Ne/H. Red plus : starbursts; blue cross : LINER AGN ; black dots : AGN.
}

\end{center}
\end{figure*}

We have investigated 
whether the  line ratio modelling  results  are constrained by the number and type of the observed lines, in terms of degeneracy.
We have discussed, in particular, the  Xia et al sample which do not show line ratios   
other than  [OIII]5007+/\Hb and [OII]3727+/\Hb.
The models  are  hardly constrained.
To  investigate the results found in Sect. 3  about the  leading   role of
the [OIII] 4363 /\Hb line ratio, we presented the following test.
A first  grid of models  was run adopting O/H ratios  as close to solar (Allen 1976) as  those adopted  to fit the sample of
 galaxies  which appear  in Paper I and in the  Kakazu et al. sample. The O/H relative abundances were
 readjusted in order to best fit the data  in tune with the other input parameters.
The results appear in Table 4. Then, we recalculated
the spectrum of the Xia et al ID 195 galaxy  by O/H= 10$^{-4}$ in agreement with the results
obtained by the  methods adopted by the other authors.
The observed [OIII]5007+/\Hb and [OII] 3727+/\Hb  ratios were reproduced by both  high and low O/H models.
 To remove degeneracy we invoked the observed weakness of the [OIII] 4363 line.
We  conclude that spectra referring only to  [OIII]5007+/\Hb and [OII] 3727+/\Hb cannot be modelled
with precision  without any further indication about e.g. the FWHM of the line profiles
or the intensity of  some other lines.

\subsection{Results for different N/H}

Let us now investigate the role of the abundances of elements  different from oxygen.
Actually, the Kakazu et al and Xia et al samples
 were chosen by our investigation because, due to the observing difficulties at those z, the lines corresponding
to heavy elements, other than oxygen,  were not reported. Those lines are generally weaker than the  oxygen ones,
and/or they cannot be easily deblended.
Nevertheless, the gas is generally  composed by the most prominent elements.
We have used in the  present models  Allen (1976) solar abundances 
(He/H=0.1, C/H=3.3 10$^{-4}$, N/H=9.1 10$^{-5}$, O/H=6.6 10$^{-4}$,
Ne/H=1. 10$^{-4}$, Mg/H=2.6 10$^{-5}$, Si/H=3.3 10$^{-5}$, S/H=1.6 10$^{-5}$, Cl=4. 10$^{-7}$, 
A/H=3.3 10$^{-6}$, Fe/H=3.2 10$^{-5}$).

In Fig. 5 we  have added to the modelling results calculated in Paper I   the results  calculated 
for the galaxy samples reported in the present work.
The  relatively large galaxy sample presented in Paper I,  contains galaxies which were selected  among those 
including at least the [NII]6548+6584 doublet, in order to minimize degeneracy.
The O/H relative abundances presented in Tables 2, 4 and 5 follow the trend found for different types of galaxies 
(Fig. 5, left middle diagram).

Changing the  abundance of one of the heavy elements relative to H, in particular for strong coolants   e.g. O, N, etc,
the results of  the emitted line ratios would  change. 
In fact, the relative abundance of each element affects not only the line intensity but also
the cooling rate of the gas in the recombination zone.

Let us investigate whether our models which adopt solar abundances for the elements corresponding
to unobserved  lines  lead to trustful results.
This is a critical test which could invalidate  most of  the results  presented in this paper.
In Paper I we  have obtained by modelling the spectra reliable N/H and O/H for each galaxy. 
The N/H versus O/H relative abundances are shown in Fig. 6.
Even with a large scattering,  the data show that there is an increasing trend of N/H with O/H.
So in order to investigate  degeneracy  which   may  result by  changing  the  N/H input in a spectrum,
we have run models with N/H lower than solar  for  galaxies which show a relatively low O/H.

We report in Table 6 the oxygen line ratios observed and calculated for two galaxies : ID 9 
(NB912)  from the Kakazu et al sample and ID 339 from the Xia et al sample. Both refer to a relatively low O/H 
(Tables 2 and 4).
Table 6 shows that the N/H  relative abundance does not affect the results  as much as to imply   a new
set of the other  input parameters.
The observational error  is $<$ 10\% for  most of the lines. 
On the other hand, changing the parameters such as \n0,  \Vs, and/or $F$ can lead to strong differences in the 
(oxygen) line ratios (see Contini \& Viegas 2001a,b).
Therefore,  the results of the present work  are  safe   
 concerning the physical conditions and the O/H relative abundance. 

As  confirmed by  Fig. 6, different
N/H  can correspond to the solar O/H for many galaxies.
 The  different  N and O trends   can be understood following Edmunds \& Pagel (1978),
namely, by the distinction between 'primary' elements such as oxygen and the   'secondary' ones.
 Nitrogen is a secondary element contaminated, however, by a non negligible primary component.
Edmunds \& Pagel  suggest that {\it although  oxygen is instantaneously recycled by the supernova synthesis, 
nitrogen could be released by longer-lived stars and hence will appear in the ISM with a delay.}

In other words, the scattering of the N/O abundances at a relatively constant O/H which appears at z$\leq$ 0.1 can be
 explained by  a  retarded release
of N  produced in low mass longer-lived stars compared with O produced in massive,
short-lived stars (Mouhcine \& Contini 2002).

\begin{table}
\centering
\caption{Comparison of line ratios to \Hb=1 calculated  with  different N/H}
\tiny{
\begin{tabular}{lccccccccc} \hline  \hline
ID     &   z  & [OIII]5007+  &  [OIII]4363 & [OII]3727+ &  O/H &       N/H  \\ 
         &      &              &             &            & 10$^{-4}$ & 10$^{-5}$ \\ \hline
9 $^1$ & 0.833& 5.97         & $<$0.12     & 1.57       & -    &  -   \\
m1     &      & 6.0          & 0.037       & 1.62        & 2.&10.  \\
m2     &      & 6.1          & 0.038       & 1.63       & 2.&6.  \\
m3     &      & 6.16         & 0.039       & 1.65       & 2.&2.  \\
m4     &      & 6.18         & 0.04        & 1.66       & 2.&1.   \\

339 $^2$ &0.602& 5.07       &  -           & 1.38      &   -   &   -    \\
m11      &     &5.42         &  -           & 1.30        & 1.8 & 10.  \\
m12      &     &5.51        &  -           & 1.31       & 1.8 & 6. \\
m13     &      &5.54        &  -           & 1.33       & 1.8 & 2. \\
m14     &      & 5.55       &  -           & 1.34       & 1.8 & 1. \\ \hline
\end{tabular}}

 $^1$ from the Kakazu et al. sample ; $^2$ from the Xia et al. sample

\end{table}

\begin{table}
\centering
\caption{Comparison of model calculated O/H with Kakazu et al results}
\begin{tabular}{ccccccc} \hline  \hline
\  ID  &   z  & mod   & KCH1$^1$ & KCH2 $^1$    \\ \hline 
\ NB816&      &       &      &        \\
\ 40   &0.629 &   8.82 &7.86 $<$ 8.03$<$8.25 & 7.84$\pm$ 0.02 \\ 
\ 76   &0.6319&  8.82  & $>$8.55& 7.89$\pm$0.01  \\
\ 118  &0.6439&  8.78  & 6.93$<$7.16$<$7.44   & 7.67$\pm$0.03\\
\ 195  &0.628 & 8.48   & 6.78$<$7.06$<$7.44   & 7.53$\pm$0.03\\
\ 252  &0.64  & 8.81   & 7.68$<$7.87$<$8.14   & 7.72$\pm$0.01 \\
\ NB912&      &         \\
\ 3    &0.82   & 8.78  &7.26$<$ 7.43$<$7.65   & 7.71 $\pm$0.02 \\
\ 6    &0.83  &  8.62  & 7.40$<$7.68$<$8.14   & 7.79$\pm$0.03 \\
\ 9    &0.833 &  8.30  & $>$7.7 & 7.72$\pm$0.02 \\
\ 10   &0.829 &  8.82  &7.59$<$ 7.72$<$7.89   & 7.74$\pm$0.02 \\
\ 20   &0.820 &  8.78  &7.18$<$ 7.36$<$7.58   & 7.55$\pm$0.02  \\
\ 239  &0.8273&  8.81  & $>$7.34&7.43$\pm$0.02 \\
\ 270  &0.8176&  8.81  & 7.28$<$7.43$<$7.61   & 7.50$\pm$0.02 \\
\ 60 $^2$ &0.393  & 8.82 &7.67$<$ 7.96$<$8.57 & 7.81$\pm$0.02 \\ \hline
\end{tabular}

$^1$ Kakazu et al (2007, tables 4 and 5). KCH1:data obtained by direct method, KCH2:data based on Yin et al (2007) method. 


\caption{Comparison of model calculated O/H with Xia et al results}
\centering
\begin{tabular}{ccccccc} \hline  \hline
\ ID   &   z     & mod&  Xia et al $^1$  &   \\ \hline
\ 339  & 0.602&  8.25&  8.10$^{+0.20}_{-0.16}$ &   \\
\ 364  & 0.642&  8.30&  8.22$^{+0.16}_{-0.13}$  &  \\
\ 246  & 0.696&  8.82&  7.71$^{+0.28}_{-0.27}$   &    \\
\ 454  & 0.847&  8.82&  8.25$^{+0.23}_{-0.23}$   &  \\
\ 258  & 0.998&  8.75&   7.49$^{+}_{-0.17}$  & \\
\ 432  & 1.573&  8.79&  8.25$^{+}_{-0.26}$   &  \\
\ 563  & 1.673&  8.79&  8.37$^{+}_{-0.28}$   &  \\
\ 103  & 1.682&  8.79&   7.97$^{+}_{-0.22}$  &  \\
\ 195  & 1.745&  8.7 &  8.38$^{+}_{-0.27}$   & \\
\ 242  & 2.070&  8.66&  8.32$^{+}_{-0.29}$   &   \\
\ 578  & 2.315&  8.66&  8.27$^{+}_{-0.26}$   &    \\ \hline

\end{tabular}

 $^1$ Xia et al (2012, table 2).  Results obtained by R23  empirical calibrators.

\caption{Comparison of model calculated O/H with Henry  et al results}
\centering
\begin{tabular}{cccccccc} \hline  \hline
\ ID   &   z      & mod$_{SB}$$^1$& mod$_{AGN}$$^2$ & KK04$^3$ & Maiolino$^4$ & FMR$^5$  \\ \hline
\ 1    & 1.82     &8.82  & 8.6   & 8.16$^{+0.16}_{-0.15}$ & 7.59$^{+0.44}_{-0.16}$  & -     \\
\ 2    & 1.73     &8.82  & 8.78  & 8.43$^{+0.16}_{-0.11}$ & 8.03$^{+0.33}_{-0.33}$  & 8.17    \\
\ 3    & 1.77     &8.82  & 8.82  & 8.65$^{+0.14}_{-0.11}$ & 8.47$^{+0.14}_{-0.26}$  & 8.33   \\
\ 4    & 1.74     &8.82  & 8.78  & 8.82$^{+0.08}_{-0.10}$ & 8.68$^{+0.09}_{-0.12}$  & 8.60    \\ \hline

\end{tabular}

\flushleft

$^1$ calculated by models referring to the SB;

$^2$ calculated by models referring to the AGN

$^3$ calculated by the  Kobulnicky \& Kewley (2004) calibration; 

$^4$ calculated by the Maiolino et al. (2008)  calibration;

$^5$ the metallicity predicted from the local FMR (Mannucci et al 2012)

\end{table}

\begin{figure}
\includegraphics[width=9.0cm]{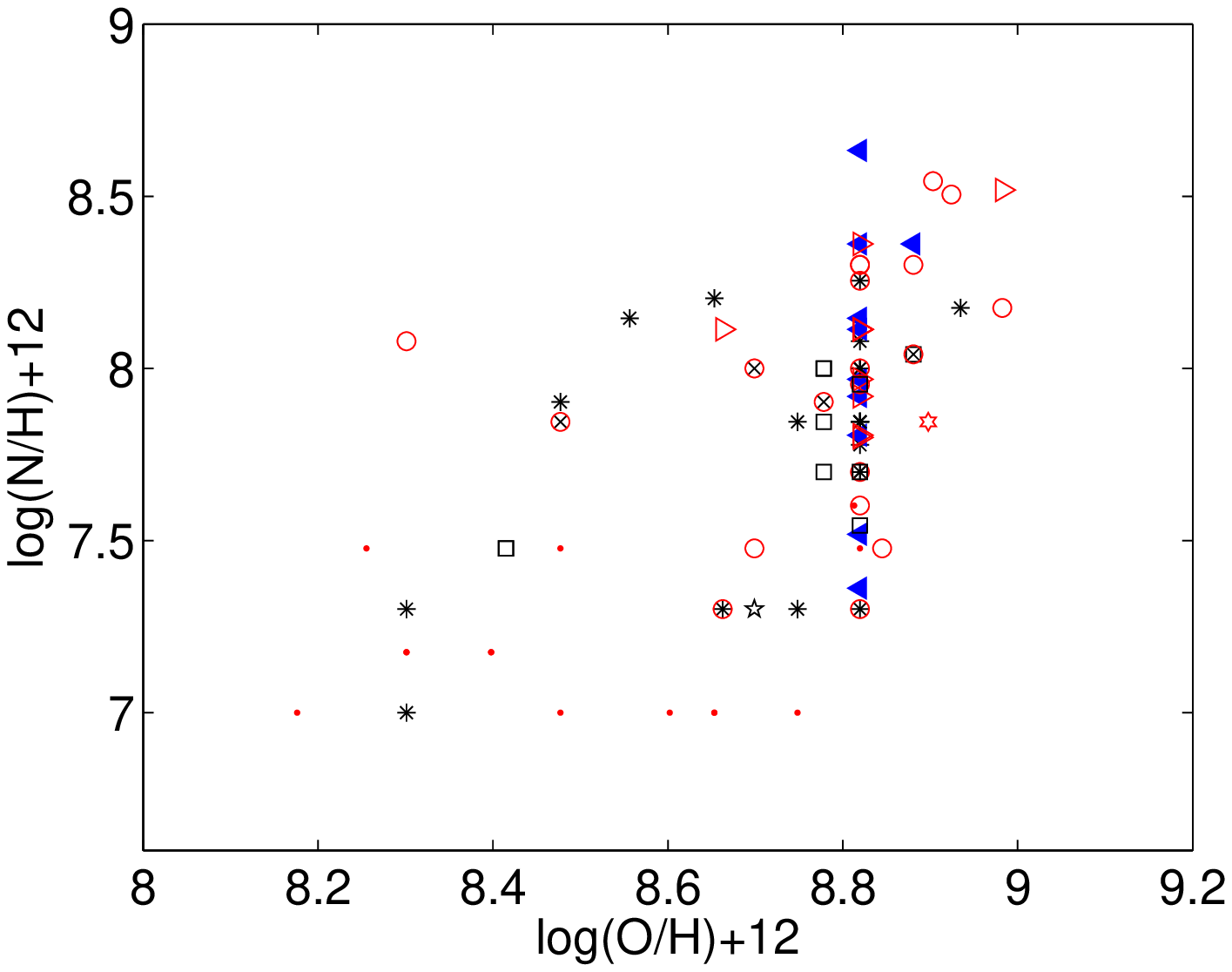}
\caption{N/H versus O/H  from the sample of galaxies presented in Fig. 5.
}

\includegraphics[width=9.0cm]{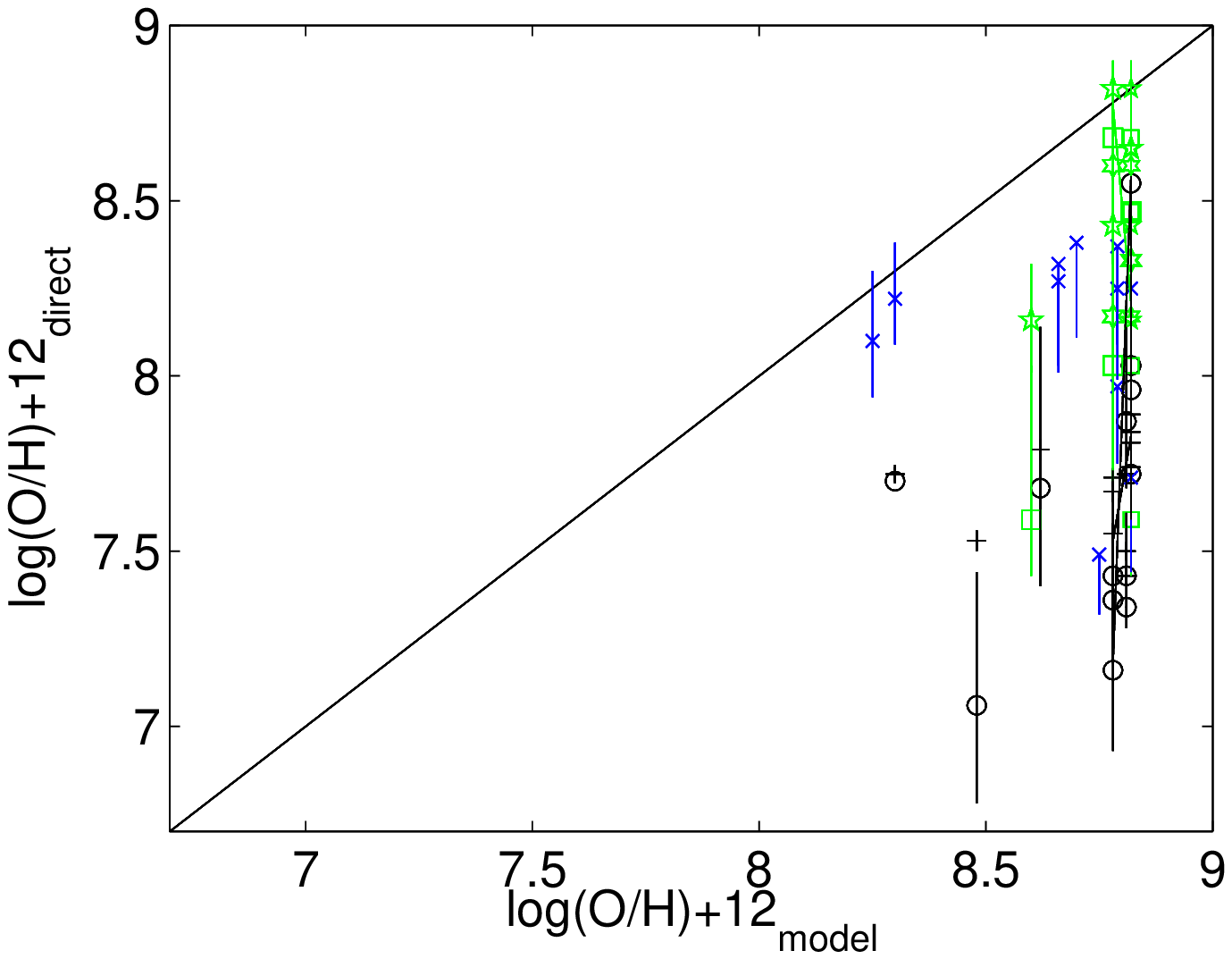}
\caption{O/H calculated by the direct method   and other empirical methods
versus O/H calculated by detailed modelling.
Kakazu et al  data : black circles (direct method);
black +  (Yin et al method);  Xia et al data : blue crosses ; green symbols
refer to the Henry et al data: large symbols result from AGN models The solid line 
represent the 1:1 line.
}
\includegraphics[width=9.0cm]{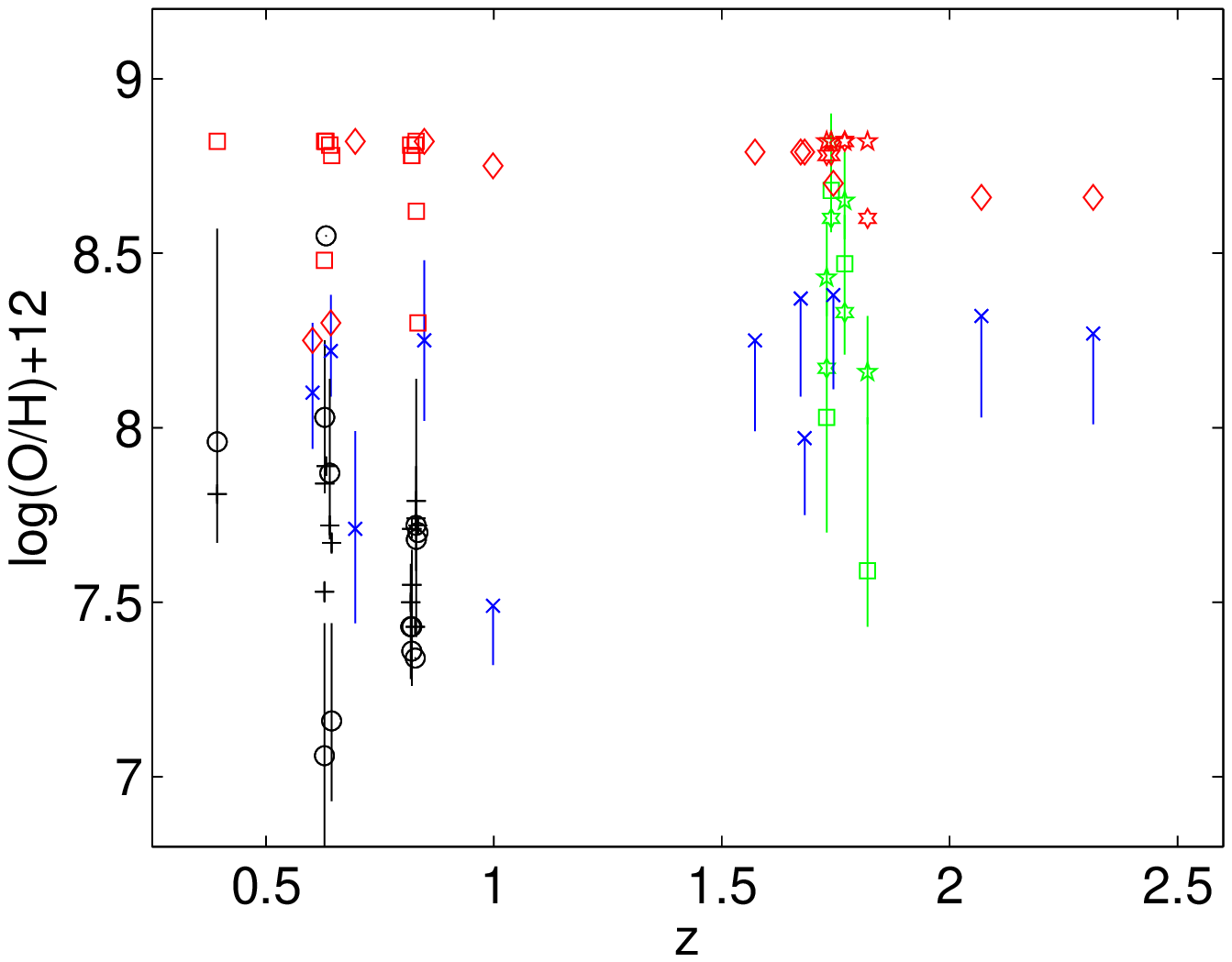}
\caption{The O/H relative abundances as a function of z.
Kakazu et al   : black circles (direct method);
black +  (Yin et al method); red squares (models). Xia et al data ; blue crosses; 
red diamonds (models);  Henry et al data (green symbols), red stars (models)
}
\end{figure}

\subsection{Comparison of the O/H results}
Finally,
we   compare in Tables 7, 8 and 9 the results for O/H calculated by the present  models and those determined
 by Kakazu et al., adopting the direct method and   by  Xia et al. and Henry et al.
adopting  empirical calibrators.
  Our results  lead to O/H = 10$^{-4}$ -- 6.6 10$^{-4}$.

 Actually, the physical parameters adopted by Kakazu et al and Xia et al. to characterize the emitting gas
are different from those determined by the best fit of the line ratios in the present paper, e.g.
Kakazu et al  fixed the electron density  to \ne=100 \cm3,  because the line ratios, which depend  directly
on the density, were not observed.
In our models \ne ranges between $\sim$ 600 and $>$ 1000 \cm3   due to compression  (Fig. 2) and recombination downstream.

Moreover, Figs. 4  shows that  SB star  temperatures calculated for the Henry et al sample   are higher than those
calculated from the Xia et al sample  and  higher than those calculated 
for a large sample of galaxies in Fig. 5. The ionization parameters calculated for  these stacked
spectra are lower than those calculated for the Xia et al sample, but they are consistently located  throughout the large
sample presented in Fig. 5.

Figs. 7 and 8 show that a  large gap appears   between the O/H results obtained by the direct method (and that of Yin et al 2007)
and those obtained by detailed calculations, however, the trend of O/H with z is roughly similar for our results and those of Xia
et al for  galaxies at z$\geq$ 1.

The large gap between our O/H results  and those  obtained by the direct method  or by empirical calibrators
has been explained in Sect. 2. Namely, the line intensities are calculated throughout a cloud integrating
on  regions showing different \Te,  \ne and, consequently, different fractional abundances of the ions.
Regions corresponding to relatively low \Te  contribute mostly to low ionization  level lines,
while a high \Te contributes to relatively high ionization level lines. So the final [OIII] and [OII] line
calculated intensities will be lower than those calculated  adopting the  optimum \Te and \ne.
To reproduce the observed [OIII]/\Hb line ratios  a higher O/H is then needed by the model.

 Concluding, the relative O/H abundances  calculated by the direct method by Kakazu et al (2007) 
and empirical methods 
by Xia et al. (2012) and Henry et al. (2013) are  lower limits because they  adopted  physical
conditions in the emitting nebulae  different from those consistently calculated by the detailed modelling.

The relatively high  O/H ratios calculated in this paper  reduce the low-metallicity  character of galaxies 
at higher z.
Moreover, Figs. 4, 5 and 8   confirm  that  the critical  redshift for the scattering of   metallicity  started  at
z $\leq$ 1.

Finally,   in this paper which deals with the modelling of
 relatively high redshift galaxies  on the basis of  [OIII]/\Hb and [OII]/\Hb  
 observations,  we claim that the  [OIII]5007+/\Hb and [OII]3727+/\Hb  
 line ratios alone are not sufficient to constrain the results.

\begin{acknowledgements}
I am very grateful to the referee for  many interesting  remarks which improved the
presentation of the paper.
\end{acknowledgements}

\section*{References}

\def\ref{\par\noindent\hangindent 18pt}
\ref Allen, C.W. 1976 Astrophysical Quantities, London: Athlone (3rd edition)
\ref Anders, E., Grevesse, N. 1989, Geochimica et Cosmochimica Acta, 53, 197
\ref Asplund, M., Grevesse, N., Sauval, A.J., Scott, P. 2009, ARAA, 47, 481
\ref Brand, K., et al. 2007, ApJ, 663, 204
\ref Capetti, A., Robinson, A., Baldi, R. D., Buttiglione, S., Axon, D.J.,
 Celotti, A., Chiaberge, M.     2013 arXiv1301.5757C
\ref Contini, M. 2013a, MNRAS, 429, 242
\ref Contini, M. 2013b, MNRAS submitted (Paper I), arXiv:1310.5447
\ref Contini, M. 2009, MNRAS, 399, 1175
\ref Contini, M. 2004a, A\&A, 422, 591
\ref Contini, M. 2004b, MNRAS, 354, 675
\ref Contini, M. 1997, A\&A, 323, 71
\ref Contini, M., Viegas, S.M. 2001b, ApJS, 137, 75
\ref Contini, M., Viegas, S.M. 2001a, ApJS, 132, 211
\ref Diaz, A.I., Prieto, M. A., Wamsteker, W. 1988, A\&A, 195, 53
\ref Edmunds, M.G. \& Pagel, B.E.J. 1978, MNRAS, 185, 77
\ref Henry, A. et al 2013, arXiv:1309.4458
\ref Kakazu, Y., Cowie, L.L., Hu, E.M. 2007, ApJ, 668, 853 
\ref Kobulnicky, H., Zaritsky, D. 1999, ApJ, 511, 118
\ref Kobulnicky, H.A., Kewley, L.J. 2004, ApJ, 617,240 (KK04)
\ref Krogager, J-K. et al. 2013, arXiv:1304.4231
\ref Maiolino, R. et al. 2008 A\&A, 488, 463 
\ref Mannucci, F., Salvaterra, R., Campisi, M. A. 2011, MNRAS, 414, 1263
\ref Mouhcine, M. \& Contini, T. 2002, A\&A, 106, 114
\ref Osterbrock, D.E.  Astrophysics of Gaseous Nebulae and Active Galactic Nuclei. Mill Valley, CA:
University Science Books; 1989.
\ref Pagel, B.E.J., Simonson, E.A., Terlevich, R.J., Edmunds, M.G. 1992, MNRAS, 255, 325
\ref Perez-Montero, E., Diaz, A.I. 2005, MNRAS, 361, 1063
\ref Ramos Almeida, C., Rodr\'{i}guez Espinosa, J.M., Acosta-Pulido, J.A. Alonso-Herrero, A.,
P\'{e}rez Garc\'{i}a, A.M, Rodr\'{i}guez-Eugenio, N. 2013, MNRAS, 429, 3449
\ref Rigby, J.R., Rieke, G.H. 2004 ApJ, 606, 237
\ref Schirmer, M., Diaz, R., Holhjem, K., Levenson, N.A., Winge, C. 2013, ApJ, 763, 60
\ref Seaton, M.J. 1975, MNRAS, 170, 475
\ref Viegas, S.M., Contini, M., Contini, T. 1999, A\&A, 347, 112
\ref Winter, L.M., Lewis, K.T., Koss, M., Veilleux, S., Keeney, B., Muschotzky, R.
2010, ApJ, 710, 503
\ref Xia, L. et al.  2012, AJ, 144, 28
\ref Yin, S.Y., Liang, Y.C., Hammer, F., Brinchmann, J., Zhang, B., Deng, L.C., Flores, H. 2007, A\&A, 462, 535
2010, ApJ, 710, 503

\end{document}